\documentclass[twocolumn]{aastex631}
\usepackage{graphicx}
\usepackage{subfigure}
\usepackage{xcolor}
\usepackage{comment}

\newcommand{\MH}{\ensuremath{{[\mathrm{M}/\mathrm{H}]}}}

\newcommand{\thing}{knot~}  
\newcommand{\thingns}{knot} %
\newcommand{\things}{knots~}
\newcommand{\Thing}{Knot~}

\begin{document}

\title{The Extremely Metal Rich \Thing of Stars at the Heart of the Galaxy}

\author[0000-0003-4996-9069]{Hans-Walter Rix}
\affiliation{Max-Planck-Institut für Astronomie, Königstuhl 17, D-69117 Heidelberg, Germany}
\author[0000-0002-0572-8012]{Vedant Chandra}
\affiliation{Center for Astrophysics $\mid$ Harvard \& Smithsonian, 60 Garden St, Cambridge, MA 02138, USA}
\author[0000-0001-6761-9359]{Gail Zasowski}
\affiliation{Department of Physics and Astronomy, University of Utah, Salt Lake City, UT 84105, USA}

\author[0000-0003-1065-9274]{Annalisa Pillepich}
\affiliation{Max-Planck-Institut für Astronomie, Königstuhl 17, D-69117 Heidelberg, Germany}

\author[0000-0003-2105-0763]{Sergey Khoperskov}
\affiliation{Leibniz Institut für Astrophysik Potsdam (AIP), An der Sternwarte 16, D-14482, Potsdam, Germany}

\author[0000-0002-7539-1638]{Sofia Feltzing}
\affiliation{Lund Observatory, Department of Geology, Sölvegatan 12, SE-223 62 Lund, Sweden}

\author[0000-0002-4013-1799]{Rosemary F.G. Wyse}
\affiliation{Department of Physics and Astronomy, The Johns Hopkins University, Baltimore, MD 21218, USA}

\author[0000-0002-6411-8695]{Neige Frankel}
\affiliation{Canadian Institute for Theoretical Astrophysics, University of Toronto, 60 St. George Street, Toronto, ON M5S 3H8, Canada}

\author[0000-0003-1856-2151]{Danny Horta}
\affiliation{
Center for Computational Astrophysics, Flatiron Institute, 162 5th Ave., New York, NY 10010, USA}

\author[0000-0001-9852-1610]{Juna Kollmeier}
\affiliation{Observatories of the Carnegie Institution for Science, 813 Santa Barbara Street, Pasadena, CA 91101, USA}

\author[0000-0002-3481-9052]{Keivan Stassun}
\affiliation{Department of Physics and Astronomy, Vanderbilt University, VU Station 1807, Nashville, TN 37235, USA}

\author[0000-0001-5082-6693]{Melissa K. Ness}
\affiliation{Research School of Astronomy 
\& Astrophysics, Australian National University, Canberra ACT 2611, Australia}

\author[0000-0001-5838-5212]{Jonathan C. Bird}
\affiliation{Department of Physics and Astronomy, Vanderbilt University, VU Station 1807, Nashville, TN 37235, USA}

\author[0000-0002-1793-3689]{David Nidever}
\affiliation{Department of Physics, Montana State University, PO Box 173840, Bozeman, MT 59717-3840, USA}

\author[0000-0003-3526-5052]{Jos\'e G. Fern\'andez-Trincado}
\affiliation{Instituto de Astronom\'ia, Universidad Cat\'olica del Norte, Av. Angamos 0610, Antofagasta, Chile   }

\author[0000-0002-7662-5475]{Jo\~ao A.S. Amarante}
\affiliation{Universitat de Barcelona Institut de Ci\'encies del Cosmos: Barcelona, Catalunya, ES }

\author[0000-0003-3922-7336]{Chervin F. P. Laporte}
\affiliation{Universitat de Barcelona Institut de Ci\'encies del Cosmos: Barcelona, Catalunya, ES }

\author[0000-0001-5258-1466]{Jianhui Lian}
\affiliation{South-Western Institute for Astronomy Research, Yunnan University, Kunming, Yunnan 650091, People's Republic of China}


\newcommand{\MHXP}{\ensuremath{\mathrm{[M/H]}_{\mathrm{XP}}}}



\begin{abstract}
\noindent
We show with Gaia XP spectroscopy that extremely metal-rich stars in the Milky Way (EMR; $\mathrm{[M/H]}_{\mathrm{XP}} \gtrsim 0.5$) -- but only those -- are largely confined to a tight ``knot” at the center of the Galaxy. 
This EMR knot is round in projection, has a fairly abrupt edge near $\sim 1.5$~kpc, and is a dynamically hot system. 
This central knot also contains very metal-rich (VMR;  $+0.2\le \mathrm{[M/H]}_{\mathrm{XP}} \le +0.4$) stars. 
However, in contrast to EMR stars, the bulk of VMR stars form an extended, highly flattened distribution in the inner Galaxy ($R_{\mathrm{GC}}\lesssim 5$\,kpc). 
We draw on TNG50 simulations of Milky Way analogs for context and find that compact, metal-rich knots confined to $\lesssim 1.5$~kpc are a universal feature. 
In typical simulated analogs, the top 5-10\% most metal-rich stars are confined to a central knot; however, in our Milky Way data this fraction is only 0.1\%.  
Dust-penetrating wide-area near-infrared spectroscopy, such as SDSS-V, will be needed for a rigorous estimate of the fraction of stars in the Galactic EMR knot. 
Why in our Milky Way only EMR giants are confined to such a central knot remains to be explained. 
Remarkably, the central few kiloparsecs of the Milky Way harbor both the highest concentration of metal-poor stars (the `poor old heart') and almost all EMR stars. 
This highlights the stellar population diversity at the bottom of galactic potential wells.
\end{abstract}

\keywords{}


\section{Introduction} \label{sec:intro}

The global formation history of a galaxy is encoded in the orbit--age--elemental abundance distribution of its stars \citep[e.g.,][]{Freeman2002, Binney2010, Rix2013, Minchev2013, Hayden2015, Belokurov_disk_halo, Weinberg2019,Belokurov_splash,Xiang2022,Rix2022, Chandra2023, Horta2024}. This present-day distribution reflects the combined effects of the hierarchical agglomeration of material, the successive star formation with its entailed feedback and enrichment, and any subsequent impulsive or secular orbit changes. While age and abundance are (largely) immutable ``birth tags'' of the stars, the observable present-day orbits may greatly differ from the stars' birth orbits. Our Milky Way has long been recognized and used as a unique laboratory to observe and then interpret this distribution \citep[e.g.,][]{Gilmore1989,Rix2013,BlandHawthorn2016}. The last decade has seen both the arrival of transformational data from the Gaia mission \citep{GaiaCollab2016,GaiaCollab2023} and an exponential growth of high-quality stellar abundances from a variety of ground-based spectroscopic surveys \citep[e.g., APOGEE and GALAH;][]{Majewski2017,DeSilva2015}. These data have made it ever more clear that our Galactic orbit--age--abundance distribution is richly structured, where at least five dimensions of this distribution play an important role: the stars' angular momentum ($L_z$), orbit circularity ($\eta\equiv L_z/L_c(E)$, \citep[e.g.][]{Chandra2023}), age, metallicity, and $\alpha$-enhancement. Much attention has been paid to the old and metal-poor end of this distribution, as it encodes information about early star formation and merging of sub-components, the transition from a dynamically hot to a disc-dominated system, and the physical processes of early element enrichment \citep[e.g.,][]{Arentsen2020, Youakim2020, Naidu2020, Sestito2020, Chiti2021, BelokurovKravtsov22, Rix2022}. 

By comparison, the distribution of the most {\it metal-rich} stars in galaxies such as the Milky Way has received much less attention. Several basic facts about very metal-rich stars in our galaxy ($\rm [M/H] \gtrsim +0.2$) seem observationally established: they are flattened and preferentially within R$_{GC}\lesssim 4$~kpc \citep[e.g.,][]{Ness2016b,Lian2020,Queiroz21,Johnson2022}. And stars in the innermost galaxy with super-solar metallicities have a wide range of ages \citep{Bensby2017}.

\begin{figure*}[ht!]
  \centering
  \includegraphics[width=\textwidth]{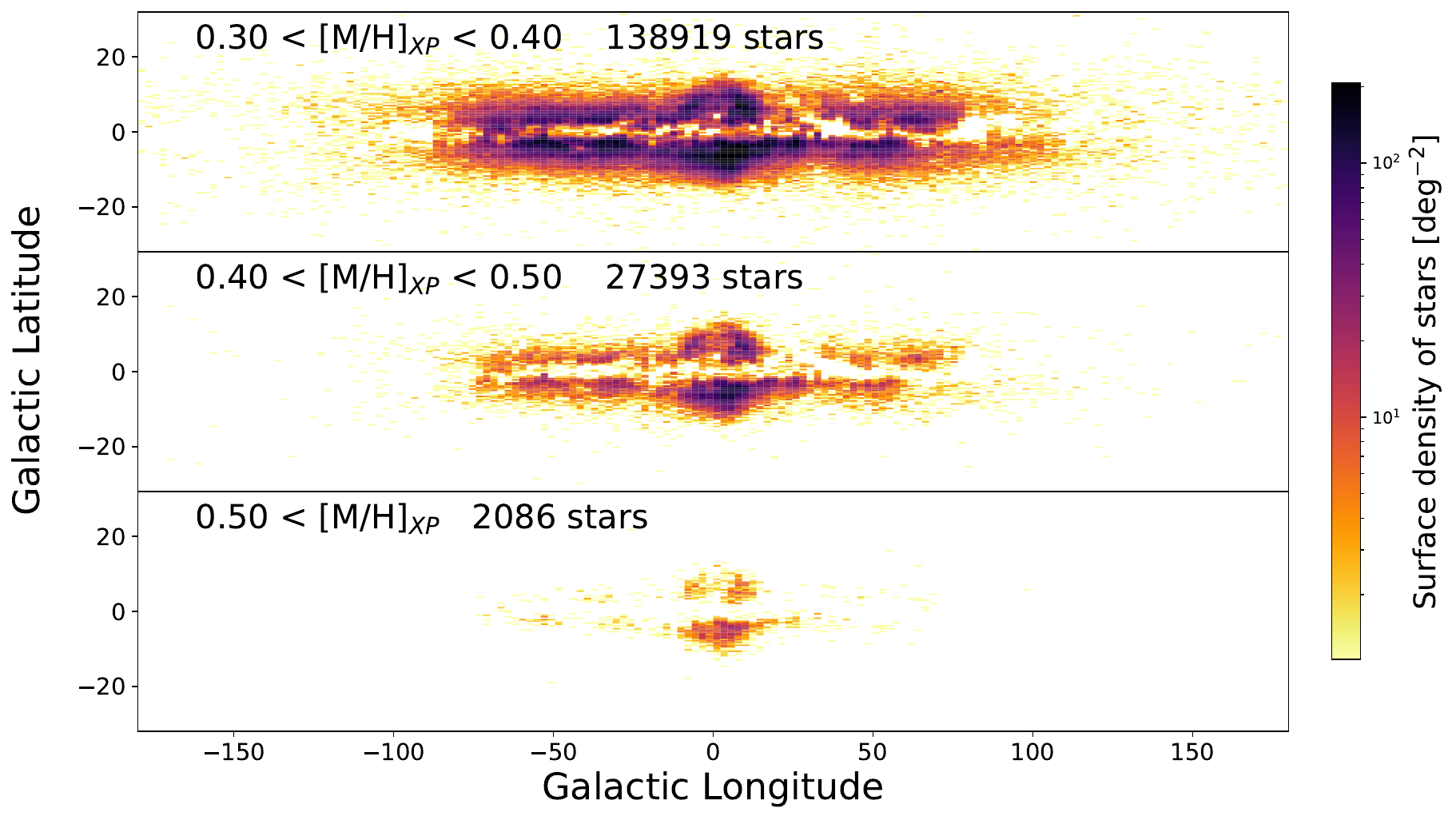}
  \caption{
   On-sky density distribution of very (and extremely) metal-rich giant stars, with $\MHXP\ge +0.3$ from Gaia DR3 XP spectra, in three different \MHXP\, bins. 
   The top panel (containing $\sim 85\%$ of all these stars) shows a flattened, disk-like distribution, concentrated towards the inner Galaxy. This central concentration increases toward higher metallicity (middle panel). 
  The bottom panel, showing the distribution of extremely metal-rich (EMR) stars with $\MHXP \gtrsim 0.5$, shows a striking change in morphology compared to the bin with $0.3<\MHXP < 0.4$: the distribution is dominated by a central \emph{\thing} with a sharp cut-off at $R_{\mathrm{GC}} \sim 1.5$~kpc. The relative prominence of this \emph{\thing} increases by an order of magnitude from the previous bin (see also Fig.~\ref{fig:EMRprofile}). Note that in all three panels, the imprint of dust extinction is apparent at the lowest latitudes; and the bar may contribute to the slight $\pm l$ asymmetry. Foreground stars closer than 4 kpc from the Sun have been removed (see sample selection criteria at the end of Section~\ref{sec:sample} }
  \label{fig:metal_rich_on_sky}
\end{figure*}

Very recent work \citep{Horta2024} points towards a central `knot'\footnote{While the 'knot' recently discovered by \citet{Horta2024} is defined somewhat differently, it is likely that there is extensive overlap between the chemo-dynamical sub-population discovered by these authors and the stars in the focus of this paper. Therefore, we adopt their terminology '\thingns' here.} of stars that are predominantly metal rich. The truly {\it most} metal-rich stars, with [M/H] almost $+1$, have only been found in the immediate vicinity ($\sim 100$~pc) of the Galactic Center \citep{Feldmeier2022}. However, absolute metallicity calibrations -- needed to compare different works -- are largely untested at $\MH>+0.5$, making such comparisons difficult \citep[e.g.][]{Soubiran2022}.  These metal-rich stars show a wide range of ages \citep[from $\lesssim 1$~Gyrs to $\sim$10~Gyr; e.g.,][]{Bernard2018,Hasselquist2020} and their detailed abundance ratios are comparable to the solar values \citep[e.g.,][]{Zasowski2019, Nepal24}. Finally, the metallicity distribution function seems to fall off very steeply beyond $\rm [M/H] > +0.3$ \citep[eg.][]{Rojas2020}, which means that these extremely metal-rich (``EMR") stars are rare. 

Qualitatively, there is a good understanding in the context of galaxy formation of what circumstances lead to very metal-rich stars. Producing such stars takes many generations of stars; it requires the ability to retain the metals produced by previous generations of stars, which is most effective deep in a gravitational potential well; and it presumably helps to be in the waning phase of an intense star formation episode, where the enrichment is still high, but there is little fresh gas supply to dilute the enrichment \citep[e.g.][]{Weinberg2017}. But the current literature is rather muted on quantitative detail: when, where, and under what {\it detailed} circumstances should EMR stars form? In addition, is there an upper limit to the stellar metallicity distribution in any galaxy; and if so, what processes or galactic properties determine the maximal metallicity?

Here, we provide an empirical starting point for understanding the creation of EMR stars in Milky Way-like galaxies
by mapping the spatial distribution of the most metal-rich stars in our Galaxy (\S\ref{sec:metal-rich-distribution}). 
The central point of this paper will be to highlight that the spatial distribution of increasingly metal-rich subpopulations
show quite a dramatic change in morphology at the extreme end (EMR stars, taken here to have $\rm [M/H] \ge +0.5$): it shifts from a predominately flattened and rotating structure to a tightly confined ($R_{\mathrm{GC}} \lesssim 1$~kpc), seemingly round \emph{``\thingns"}, with stars on radial orbits with modest net rotation.  
While this compact \thing seems dominant in the EMR regime, there are stars with similar spatial and orbit distributions also at lower metallicities, from $0<\MH <+0.6$ with a peak at $\MH\sim +0.3$.
We used the Milky Way analogs of the TNG50 simulation suite \citep{Nelson2019,Pillepich2019} to provide an initial context for the discovery of this EMR stellar \thing in the Milky Way (\S\ref{sec:galaxy_sims}). Finally, we point out avenues for follow-up work to understand this remarkable structure, in particular with SDSS-V, but also eventually with MOONS \citep{Cirasuolo2020} and 4MOST \citep{deJong2019}.

\begin{figure}[ht!]
  \centering
  \includegraphics[width=\linewidth]{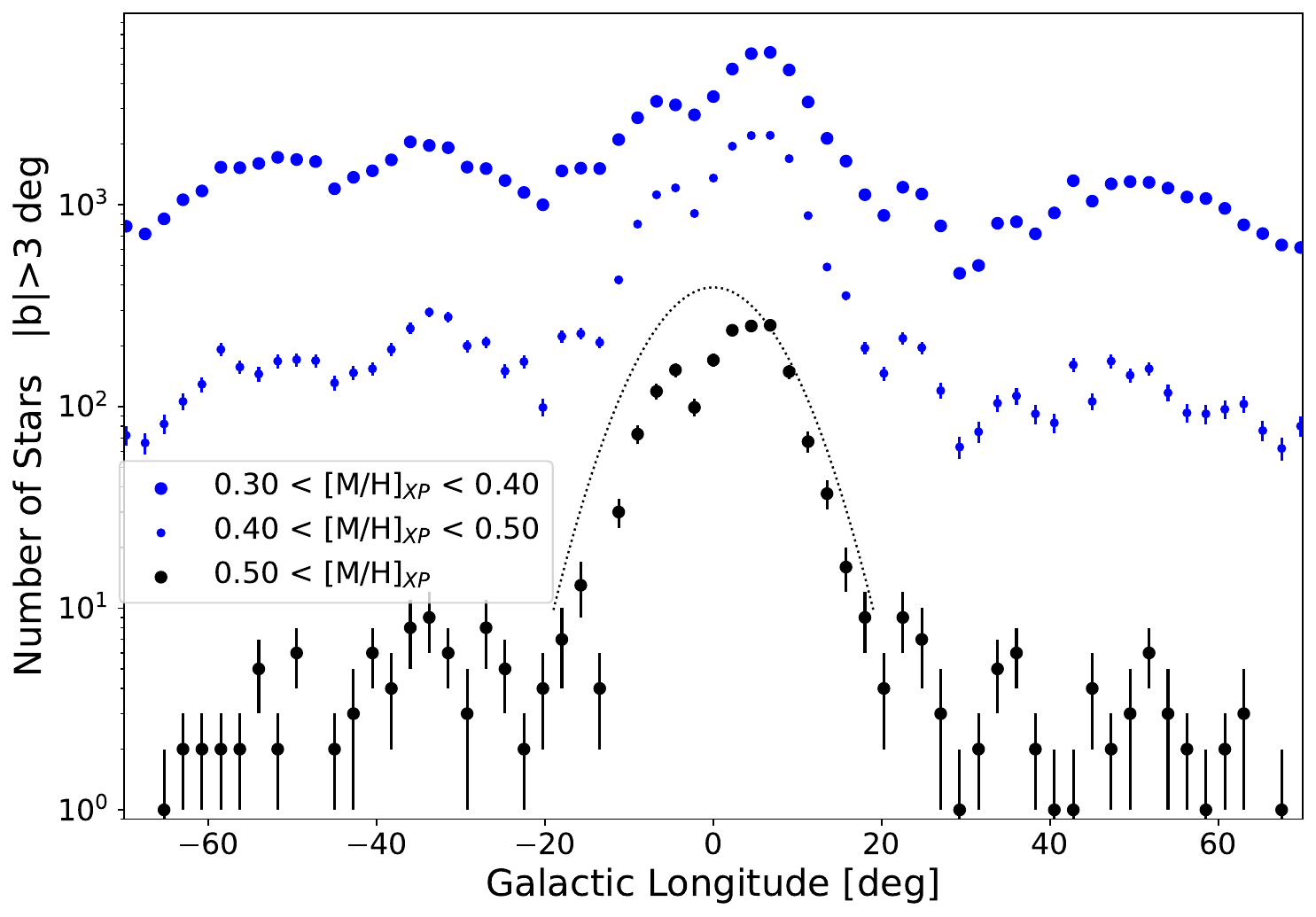}
  \caption{
   The longitudinal number density profile of the very metal-rich stars towards the center of the Milky Way. These are the same stars as shown in Fig.~\ref{fig:metal_rich_on_sky}, but integrated over Galactic latitude (excluding $\pm 3^\circ$ because of the severe dust extinction). The dashed line just illustrates a Gaussian of $\sigma=10^\circ$ centered on $|l|=0$ in order to illustrate the level of asymmetry induced by dust extinction. While for the least metal-rich bin ($0.3<\MHXP<0.4$) the density peak of the central \thing is only $\sim 3\times$ above the surrounding (in projection) disk-like distribution, this contrast has grown to $100\times$ in the EMR bin ($\MHXP > 0.5$): EMR stars are basically only found in the $\lesssim 1$~kpc \thingns.}
  \label{fig:EMRprofile}
\end{figure}

\section{The Distribution of the Most Metal-Rich Stars in the Milky Way} \label{sec:metal-rich-distribution}
\subsection{Metallicities for Giants from Gaia XP Spectra}\label{sec:sample}
To make all-sky maps of stars at a given metallicity, the Gaia~XP spectra \citep{DeAngeli2023, Andrae23a, Andrae23b} are currently the data set of choice; the higher-fidelity, all-sky spectroscopic survey SDSS-V \citep{Kollmeier2017,Almeida2023} is not yet completed. Although Gaia~XP spectra have rather low spectral resolution, they have proven to be remarkably precise and accurate in determining basic stellar parameters and abundances ($T_{\rm eff,XP}$, $\log{g_{\rm XP}}$, [M/H]$_{\rm XP}$, [$\alpha$/M]$_{\rm XP}$). This is true at least for red giant stars that have prominent metallicity-dependent spectral features, even at low resolution \citep[][]{Andrae2023, Zhang2023, Li2023}. At present, this quality of stellar labels from XP spectra has so far (only) been achieved with data-driven approaches that link these stellar labels to high-fidelity stellar surveys, such as SDSS-IV's APOGEE \citep{Majewski2017}. However, with these approaches, the precise spectrophotometry and Gaia's exceptional data consistency across the sky yield samples of precise metallicities ($\sigma_{\mathrm{[M/H]}}\lesssim 0.1$~dex) for samples of tens of millions of giants. It has also been shown that the rate of spurious [M/H]-estimates is exceptionally low, which permits the identification of rare subpopulations --- for example, \citet{Rix2022} demonstrated that metal-poor stars can be relatively cleanly identified in the inner Galaxy. 

An inherent and currently inevitable aspect of data-driven stellar labels such as \MHXP~ is that they can only be trusted in the regime that is well covered by the training set. At and beyond the edges of the stellar label regime that is well-populated by the training set, the inferred stellar labels tend to have some systematic inaccuracies but often still preserve the relative ranking of, say, the metallicities. For example, stars that are truly exceptionally metal poor are identified as such by data-driven analyses of XP spectra; however, the inferred \MHXP values are closer to the bulk [M/H] than those inferred by high-resolution analyses \citep[][Fig.4.a]{Andrae2023}.
This aspect of the data-driven \MHXP\ matters in our application at hand: we are looking for the most metal-rich stars on the sky. Given that the Gaia set is $\sim 100\times$ larger than the SDSS-IV training set, the most metal rich stars in Gaia data are at --- and slightly beyond ---  the training-set regime.
A direct cross-validation comparison with APOGEE DR17 shows that for stars with [M/H]$_{\rm APOGEE}>+0.27$, 
the scatter is 0.07~dex, but \MHXP\ metallicities are systematically underestimated by 0.04~dex at the EMR end ($\rm [M/H] \ge +0.5$). Therefore, one can trust the relative [M/H] ranking (which can identify e.g. the 'most metal-rich stars'), but should take the accuracy of the actual \MHXP\ values with grains of salt.

Here, we simply adopt the \MHXP\ values from \citet{Andrae2023} for a data-quality-cleaned RGB sub-sample and explore the extremely metal-rich end of the sample's metallicity distribution. The following data quality cuts were made here : $G\le16$~mag, $\delta\varpi/\varpi<1/4$,  T$_{eff}\le$ 5200~K; log$g<3.5$. We also eliminated foreground stars, closer than 4~kpc from the Sun, by requiring $\varpi+\delta\varpi < 0.25$. All subsequent plots and analyses start with this underlying sample.

\subsection{On-Sky Distribution of Very Metal-Rich Stars in the Galaxy}

For the stars in the sample defined at the end of Section~\ref{sec:sample} (e.g. gianst with $D_\odot > 4$~kpc), Figure~\ref{fig:metal_rich_on_sky} shows a set of three maps of very metal-rich stars ($\MHXP\ge +0.3$) across the Galactic low-latitude sky that highlight our basic observational result. 
Each of the three panels shows the density of RGB stars in bins of increasing \MHXP. The top panel, showing the most metal-poor subset of our sample (Sec.~\ref{sec:sample}) still constitutes the $\sim 2.5$\% most metal-rich members of the entire Gaia sample of giants with XP metallicities from \citet{Andrae2023}. These already very metal-rich stars are concentrated towards the inner Galaxy in a thick disk-like configuration; the details of the morphology and structure in the very midplane are of course obfuscated by the dust extinction.  
The next higher metallicity bin (middle panel; $0.4<\MHXP<0.5$) shows a qualitatively similar picture, with important differences apparent: the concentration towards the inner Galaxy is more manifest, and there is clear evidence for a central concentration, or \thingns, that appears round in projection. 
In continuing these trends, the bin of EMR stars ($\MHXP\gtrsim +0.5$) shows an on-sky distribution that appears strikingly different: it is fully dominated by a centrally confined round \emph{\thingns}, with only a small minority of stars in a disk-like structure.  This EMR core seems quite sharply confined within the central $10^\circ$, corresponding to 1.4~kpc at the distance of the Galactic center. We focus here on the angular (rather than 3D) distribution of these stars to constrain the Galactocentric extent, as many parallaxes of the sample stars have large ($\sim 20\%$) uncertainties.

\begin{figure}[ht!]
    \centering
    \includegraphics[width=\linewidth]{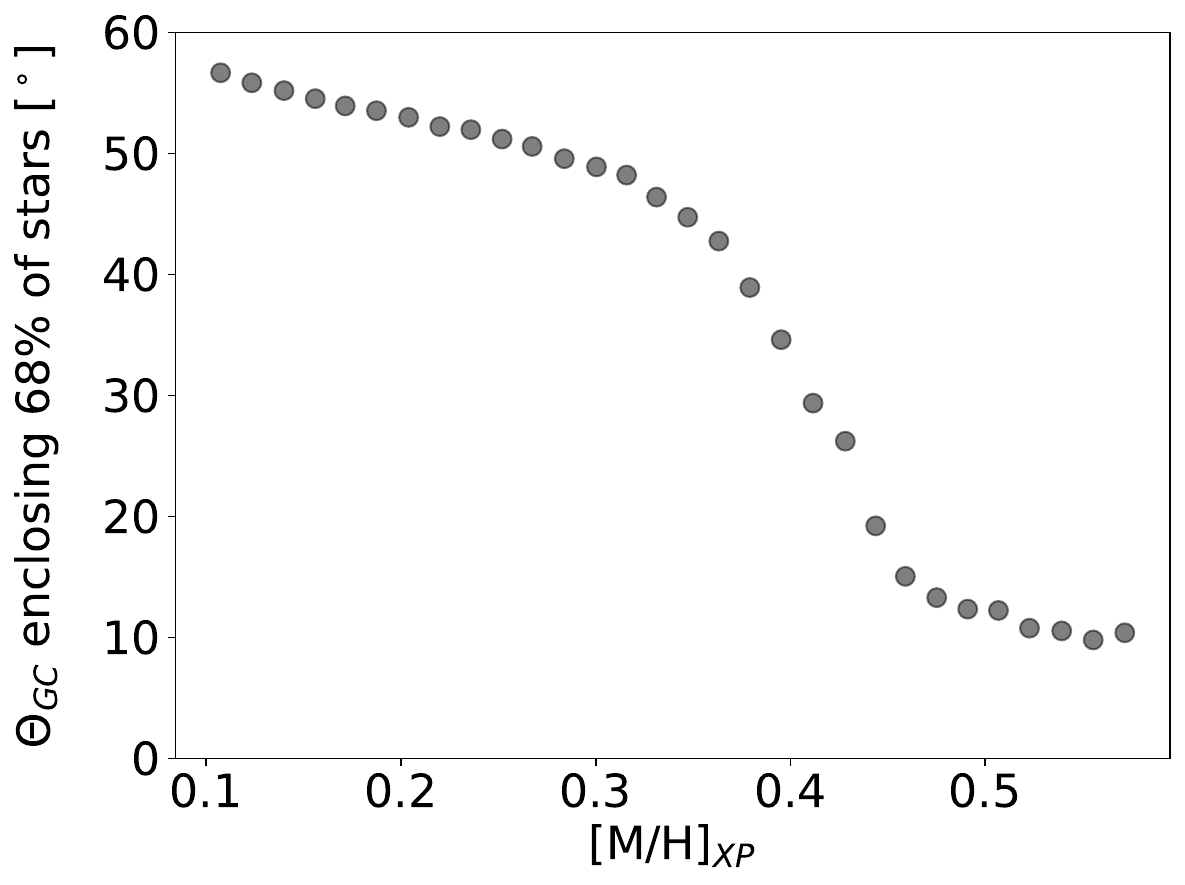}
    \caption{Central concentration of different mono-abundance population in the inner Milky Way, shown as the angular distance from the Galactic center, $\Theta_{GC}$ that includes the bulk (68\%) of the stars, as a function of \MHXP. At the Galactic center, $10^\circ$ correspond to a projected distance of 1.5~kpc. The dramatic change in appearance seen in Fig.~\ref{fig:metal_rich_on_sky} is reflected here as a large drop of $\Theta_{GC}(\MHXP)$ that is rather abrupt given \MHXP~uncertainties. The initial sample defined at the end of Section~\ref{sec:sample} was used, but only stars within $90^\circ$ of the Galactic Center were considered. Obviously, dust extinction affects this distribution in complex ways, but its effect should be similar for populations of different \MHXP .}
    \label{fig:D68percentile_MH} 
\end{figure}

As shown in Figure~\ref{fig:metal_rich_on_sky}, this dramatic change in \thingns-to-disky contrast can be quantified by considering the sample density along Galactic longitude after integrating over Galactic latitude (after we have excluded the regime $|b|<3^\circ$ where dust extinction is prohibitive). In the EMR stars (bottom panel of Figure~\ref{fig:metal_rich_on_sky}), the peak near $l\sim 0^\circ$ is nearly 100$\times$ higher than the density at $l\pm 20^\circ$ (Fig.~\ref{fig:EMRprofile}). 
This \thingns-to-disky contrast is almost an order of magnitude higher than seen in the adjacent metallicity bin ($0.4<\MHXP<0.5$). 

In interpreting the bottom panel, it must be kept in mind that the \MHXP~precision is comparable to the separation of the bins, which implies that some fraction of the stars in the bottom panel, including those more than 10$^\circ$ from the Galactic Center, are expected to be lower-metallicity contaminants.
However, taken together, Figures~\ref{fig:metal_rich_on_sky} and \ref{fig:EMRprofile} imply that the most metal-rich stars (comprising 0.035\% of the sample) are found almost exclusively in a tightly confined ($R_{\mathrm{GC}} \lesssim 1.4$~kpc) EMR stellar \thingns. 

Figure~\ref{fig:D68percentile_MH} represents this metallicity-dependent change in structure in a different way that is useful for comparison with simulations. It shows the angular distance from the Galactic Center, $\Theta_{GC}$, that encloses 68\% of the stars in different mono-\MHXP~populations. At $\MHXP\approx +0.45$ this 68\% enclosing angle drops by almost a factor of 4, from 45$^\circ$ to just above $10^\circ$. This drop is basically as abrupt as can be expected in the presence of finite \MHXP uncertainties.

\begin{figure}[ht!]
    \centering
    \includegraphics[width=\linewidth]{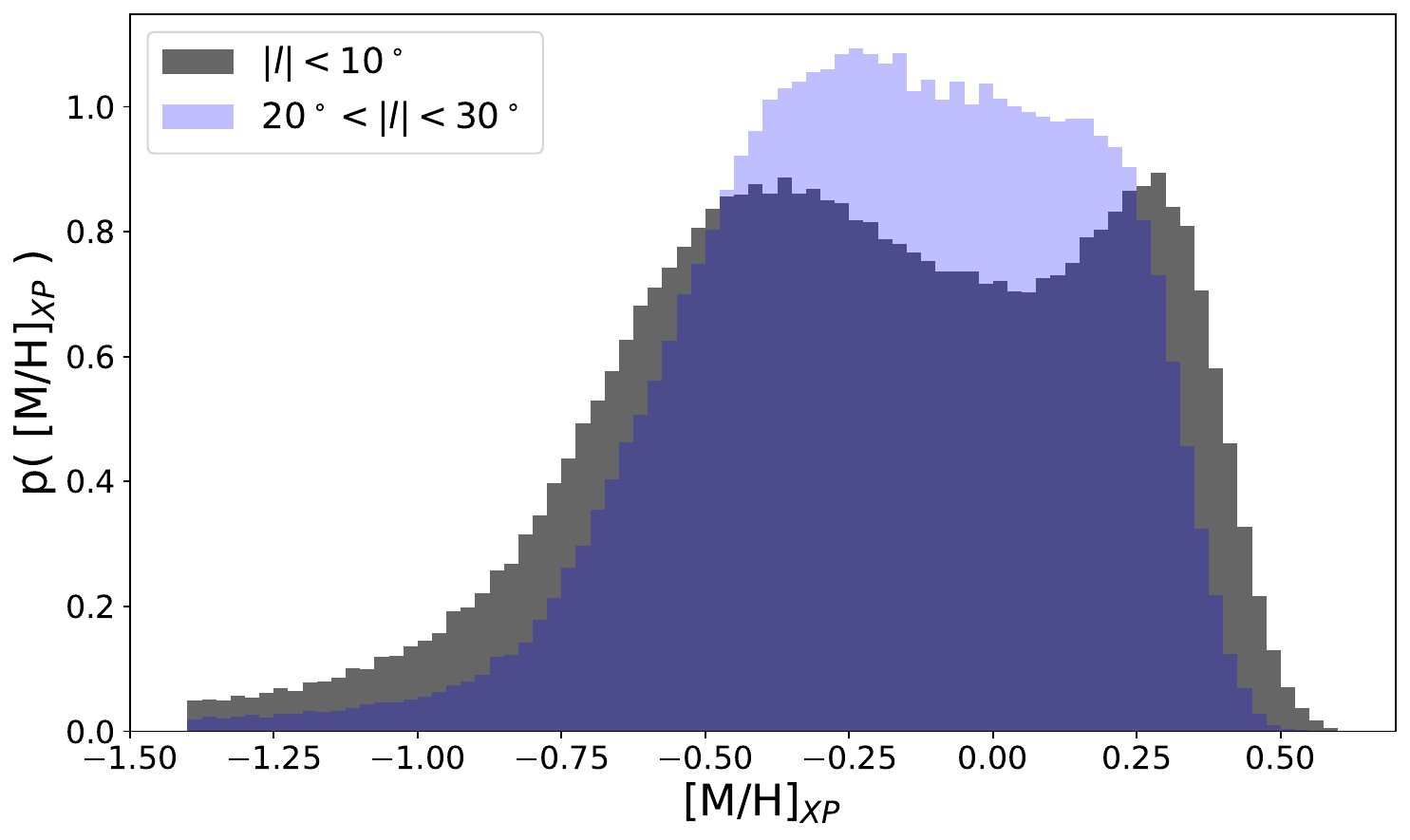}
    \caption{Metallicity distribution function (MDF) of stars within $|l|,|b|<10^\circ$ of the Galactic center (black), compared to those $20^\circ$ further out in Galactic longitude (lavender). The central MDF shows a distinct peak of very metal rich stars, reflecting the presence of a tightly confined \thing in all metallicity bins of Figure~\ref{fig:metal_rich_on_sky}. The EMR stars $(\MHXP\ge+0.5)$ are exclusively present at the center. This illustrates that the central \thing does not only contain EMR stars (which are confined to it), but also other, and more, very metal rich stars $\MHXP>+0.2$, with a peak at $\MHXP=+0.35$}.
    \label{fig:pMH} 
\end{figure}

Both Figure~\ref{fig:metal_rich_on_sky} and Figure~\ref{fig:EMRprofile} also reveal that the metallicity bins $\MHXP \le +0.4$ have a portion of their stars in the same geometry of the EMR \thingns. This implies that this \thing is not made exclusively of EMR stars with $\MHXP \ge +0.5$. We explore and quantify the metallicity distribution of this metal-rich \thing by comparing the metallicity distribution of the stars within $|l,b|<10^\circ$ to the
stars somewhat further away, $20^\circ<|l|<30^\circ$ \& $|b|<10^\circ$, in Figure~\ref{fig:pMH}. This figure shows that
$p\bigl (\MHXP\bigr )$ at the very center has a distinct peak at $\MHXP=+0.35$, seen before \citep[eg.][]{Rojas2020}. Figure~\ref{fig:pMH} shows that this is not present further away from the Galactic center. This implies two things: First, EMR stars are almost exclusively confined to a central \thingns ($R_{GC}^{projcted}\lesssim 1.5$~kpc). Second, this \thing does not contain only EMR stars. Indeed, it appears to be predominantly made up of stars that are still very metal rich presumably ($\MHXP\ge0.2$), but with a peak at $\MHXP=0.35$. The bulk of these stars may be part of -- or even constitute -- the central 'knot' of stars recently discovered by \citet{Horta2024} in an orbit-space analysis of the APOGEE sample.

\begin{figure*}[ht!]
    \centering
    \begin{minipage}{.48\linewidth}
        \includegraphics[width=\linewidth]{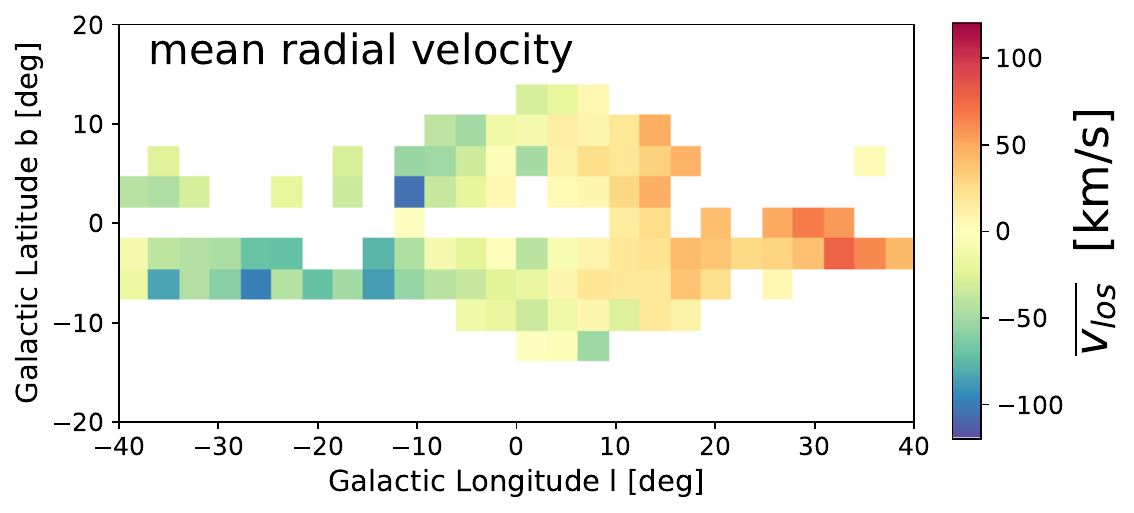}
    \end{minipage}\hfill
    \begin{minipage}{.48\linewidth}
        \includegraphics[width=\linewidth]{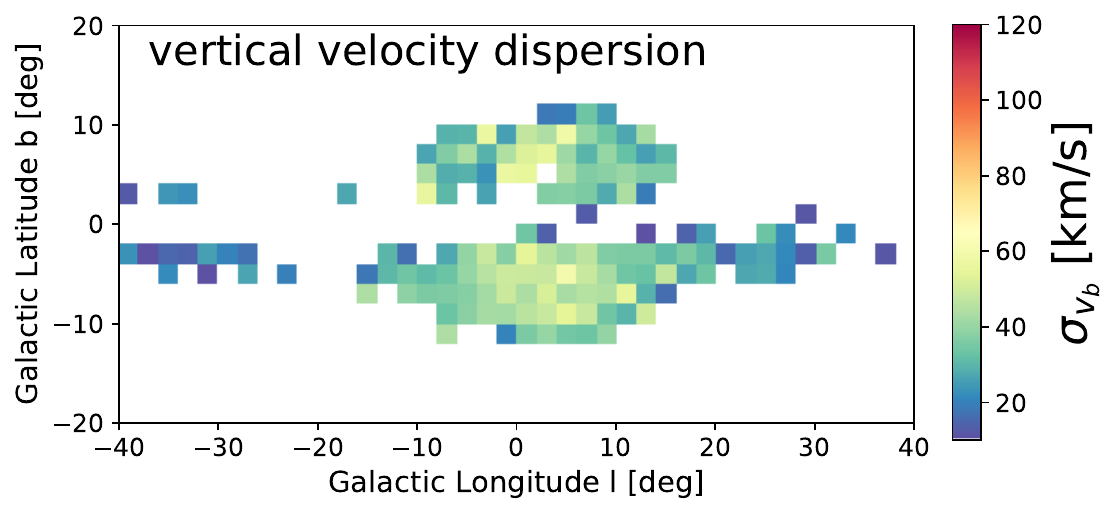}
    \end{minipage}
    
    \vspace{0.5cm}
    
    \begin{minipage}{.48\linewidth}
        \includegraphics[width=\linewidth]{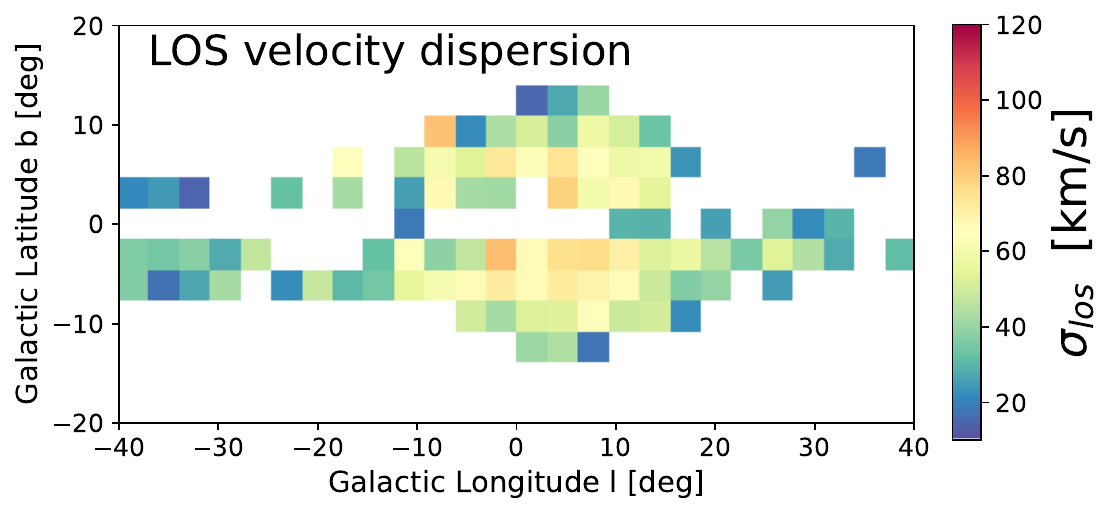}
    \end{minipage}\hfill
    \begin{minipage}{.48\linewidth}
        \includegraphics[width=\linewidth]{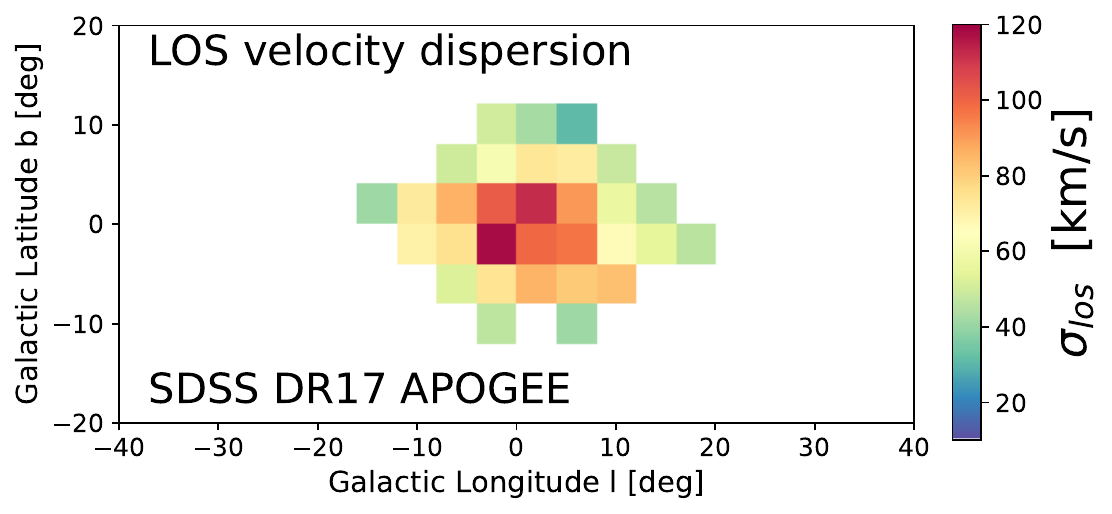}
    \end{minipage}
    
    \caption{Kinematics of the EMR \thing of the Galaxy (here taken as $\MHXP>0.45$ to slightly enlarge the sample). Top left: Mean line-of-sight (LOS) velocity. Top right: vertical velocity dispersion projected onto the sky. Bottom left: LOS velocity dispersion, of stars with radial velocities from Gaia. Bottom right: LOS dispersion from SDSS/APOGEE DR17, which also cover the cey center that is too extincted to be ``seen" by Gaia. There is modest mean rotation of 35~km/s at $|l| \sim 7^\circ$ from the Galactic center. The line-of-sight velocity dispersion rises rapidly from $\sim 75$~km/s at $\sim 7^\circ$ from the Galactic center to nearly 120~km/s towards the Galactic Center, only seen in the SDSS/APOGEE DR17 data. The \thing appears to be dynamically hot system with little rotation. The steep central rise of the LOS velocity dispersion points towards radial orbits.}
    \label{fig:EMRkinematics}
\end{figure*}

\subsection{Kinematics of the Most Metal-Rich Stars in the Galaxy}

This EMR \thingns, identified in position--\MHXP~space, deserves a thorough kinematic or orbital analysis, which is beyond the scope of this article. But some initial kinematic considerations are illuminating, as shown in Figure~\ref{fig:EMRkinematics}. A significant subset of stars with \MHXP\ have radial velocities from Gaia RVS, and hence estimates of their motions. For this kinematic analysis we consider a slightly more inclusive definition of EMR \thing ($\MH > 0.45$), to obtain a larger sample, while still maintaining a high \thingns-to-disky contrast. 

The upper left panel of Figure~\ref{fig:EMRkinematics} shows that the EMR \thing has little net rotation ($\sim 30$~km/s) at $l\pm 10^\circ$. Its vertical velocity dispersion is $\sim$60~km/s, three times higher than at longitudes $\ge 30^\circ$. The line-of-sight velocity dispersion, $\sigma_{LOS}$, at $5-10^\circ$ from the GC is $\sim 70$~km~s$^{-1}$.  From Gaia~XP data the LOS dispersion closer to the Galactic center cannot be measured because of dust. However, the dust-penetrating APOGEE spectra from SDSS-IV can do this much better, despite their uneven spatial sampling and smaller sample size. They show that the LOS velocity dispersion of these EMR stars rises steeply towards the Galactic center, reaching $\sim 120$~km/s, consistent with earlier findings by \citet{Ness2016a,Zasowski2016}. Qualitatively, such a steep increase in $\sigma_{LOS}$ is expected from systems with radial orbits \citep[e.g.][]{BT2008}. Taken together, these observations imply that the EMR \thing is a dynamically hot system with little rotation, with its stars are on seemingly radial orbits.

These results are broadly consistent with earlier measurements, such as \citet{Rojas-Arriagada14} and \citet{Valenti2018}, although those lacked either metallicity differentiation or extensive spatial coverage.

\subsection{The Age Distribution of the EMR \Thing}
\label{sec:ages}

\begin{figure}[ht!]
  \centering
  \includegraphics[width=0.5\textwidth]{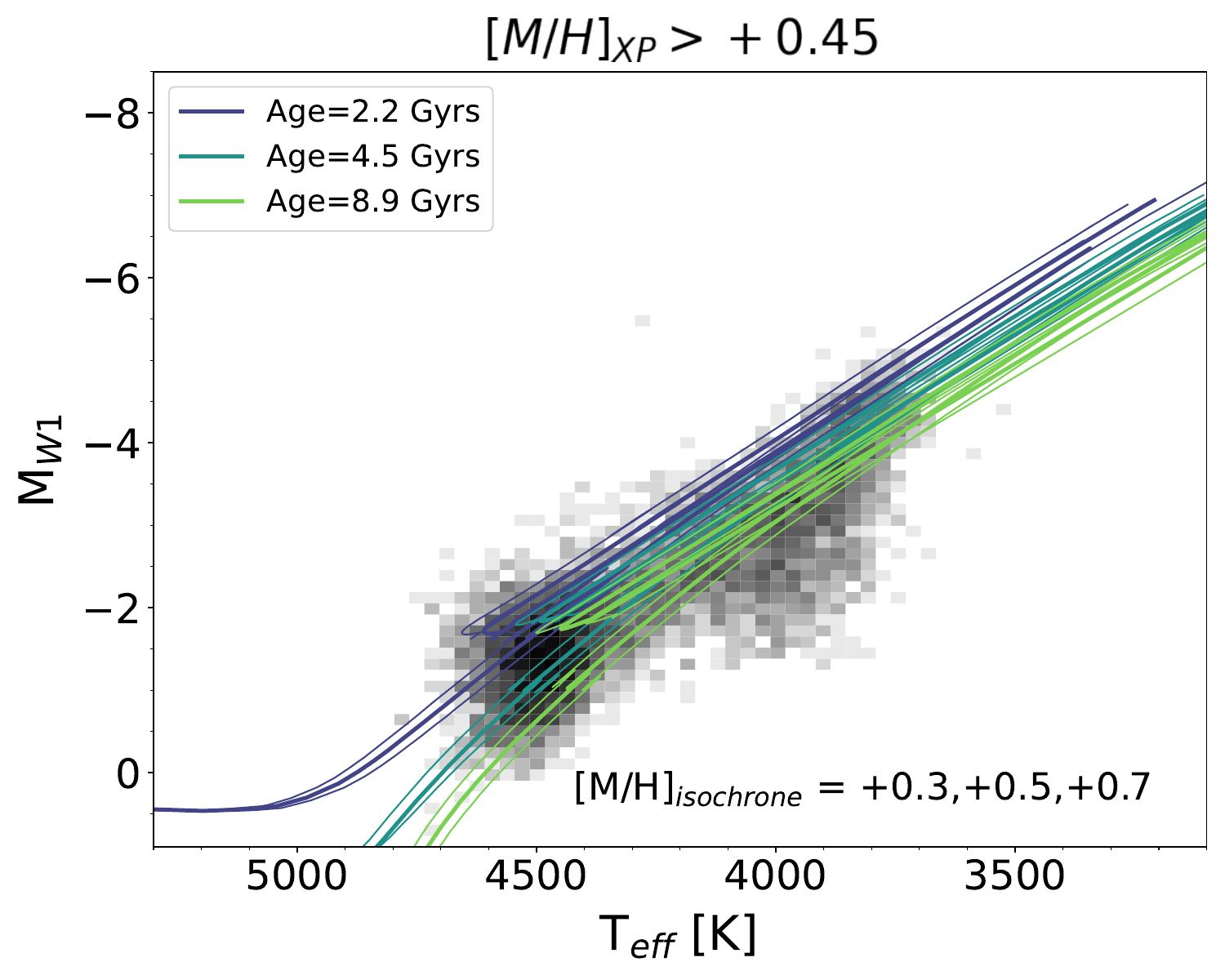}
  \caption{Color-magnitude distribution of the EMR giants (with $\MHXP>+0.45$) within 15$^\circ$ of the Galactic center, shown in grey as the (reddening-insensitive) $T_{eff,XP}$ \emph{vs.} the absolute magnitude in the WISE W1 band, $M_{W1}$.  Overlaid are three sets of Padova isochr
  ones \citep{Bressan2012} for ages of 2--9~Gyr, and for metallicities of $+0.3,+0.5$ and $+0.7$. The CMD positions implies that the stars have a considerable range of ages ($\sim 3-10$~Gyrs). More detailed fitting is precluded by the limited distance (parallax) precision, which lead to $M_{W1}$ uncertainties of $\sim 0.2$~mag; typical $T_{eff}$ uncertainties are $\lesssim 50$~K \citep{Andrae2023}.
  }
  \label{fig:RGB_in_CMD}
\end{figure}

We also briefly explore what the data indicate about the age distribution of these most metal-rich stars in the Milky Way. Figure~\ref{fig:RGB_in_CMD} shows these EMR stars with $\MHXP>+0.45$ in a reddening-insensitive version of the CMD: the absolute magnitude in the WISE~1 band versus the effective temperature of the catalog of XP spectra \citep[both from the][]{Andrae2023}. The main limitation of this CMD is the precision of the parallaxes for these stars, typically with $\delta\varpi/\varpi > 0.1$, which leads to $\sim 0.2$~mag uncertanities in $M_{W1}$. Isochrones of $\rm [M/H]=+0.3$, $+0.5$ and $+0.7$ for three different ages (2.2, 4.5, and 8.9~Gyr) are overlaid.  While the data quality does not warrant a detailed analysis, the CMD shows that these very metal-rich stars appear to be of intermediate age in terms of Galactic evolution, between 3 and 10 Gyrs. These results are very much in line with the earlier findings about the age distribution of super-solar main sequence stars by \citet{Bensby2017}.

\section{Galaxy Formation Simulations as Context}
\label{sec:galaxy_sims}

We now turn to cosmological simulations to explore to which extent this metallicity-dependent structure is expected from current simulations. In the TNG50 galaxy formation simulation, \citet{Pillepich2023} recently identified and published a set of galaxies that at $z=0$ resemble our Milky Way in terms of galaxy stellar mass, disky stellar morphology and 1 Mpc-scale environment. Among the overall set of 138 galaxies, a handful of them (with IDs 516101, 535774, 538905, 550149, 552581, and 566365) may serve as particularly suitable Milky Way analogs, in that they exhibit stellar disk scale length and scale heights similar to those inferred for our Galaxy. Many of these also resemble the Milky Way in terms of the bulge-to-disk ratio\footnote{Given that there is no consensus what exactly constitutes the ``bulge'' of our own Milky Way, it is hard to match simulated galaxies to our Galaxy in this respect. The TNG simulations that we use have a median bulge-to-disk ratio of $\sim 0.25$ \citep{Zana2022}.} \citep{Zana2022}. \citet{Boecker2023} has studied the central ($\lesssim 0.5$~kpc) stellar populations in TNG50 Milky Way analogs. They found that these stars are  metal rich, have a range of ages, and were predominately born \emph{in-situ} or migrated inward.

Which subset is 'most analogous' somewhat depends on the science question at hand \citep[see e.g.][]{Chandra2023}. For example, \citet{Semenov2024} considered a different subset among all the galaxies simulated within TNG50 to be Milky Way analogs, based on their total halo mass, kinematic properties, and present-day star formation rate, resulting in 61 galaxies. We will consider both of these TNG50 subsets here.

\begin{figure*}[ht!]
  \includegraphics[width=1.1\columnwidth]{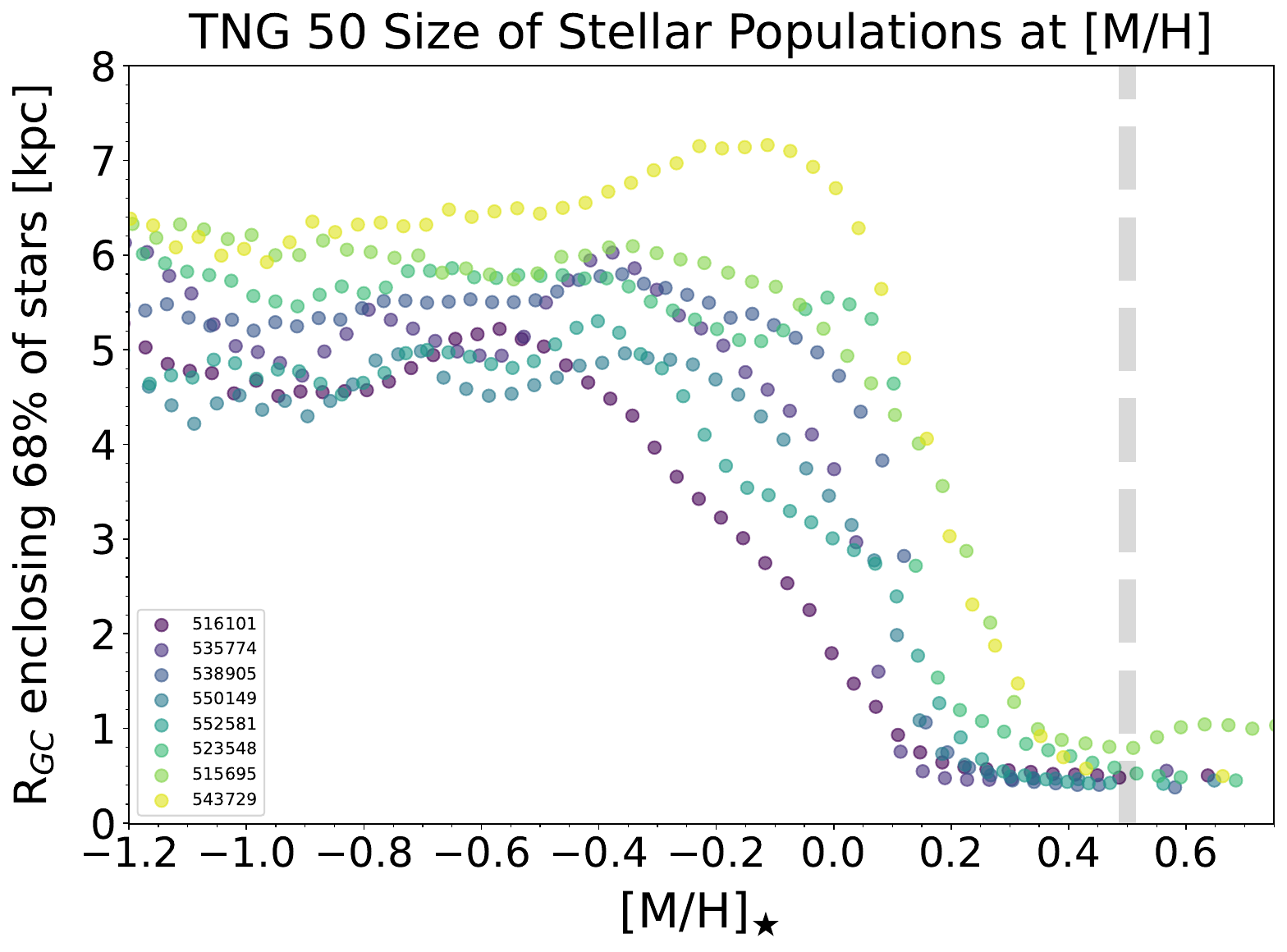}
  \includegraphics[width=1\columnwidth]{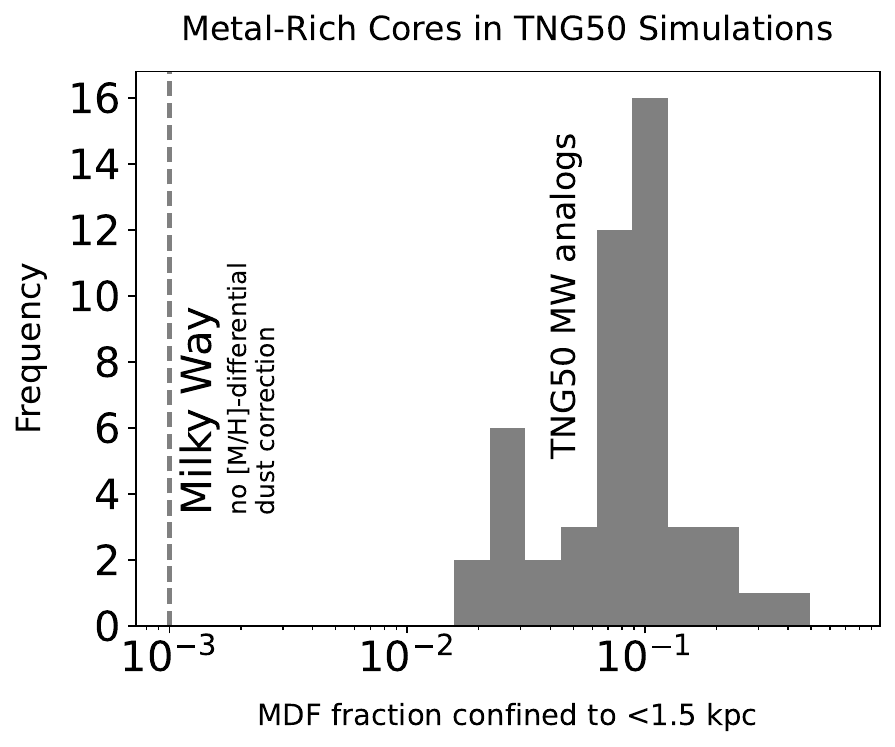}
    \caption{Radial extent of different mono-abundance stellar populations for Milky Way analogs from the TNG50 simulation \citep{Pillepich2023}. The left panel shows the radius $R_{\mathrm{GC}}$ that encloses 68\% of a mono-abundance population as a function of \MH for a handful of example galaxies. This left panel shows that for (at least these) simulated galaxies the most metal-rich stars are very centrally concentrated, forming a \thing of $\lesssim 1$~kpc. This is in qualitative agreement with our analysis of the Milky Way (see Fig.~\ref{fig:D68percentile_MH}). However, in these simulated galaxies the central confinement encompasses stars within $\Delta\MH = 0.4$ of the maximal metallicity, whereas in our Galaxy the stars in the \thing are within 0.1~dex of the maximal metallicity.  To eliminate issues revolving around metallicity systematics, we also consider which fraction of the stars' MFD is part of this central \thingns . 
    The right panel shows, for a much larger set of TNG50 Milky Way analogs, the (most metal rich) fraction of the MDF that is confined to within  $R_{\mathrm{GC}}= 1.5$~kpc: typically between 5\% and 15\% of all stars within 20 kpc. By comparison, the \emph{face value} fraction of stars in the Milky Way's \thing (i.e. in the Gaia XP sample as defined in Section~\ref{sec:metal-rich-distribution}) is dramatically lower, only 0.1\% (Fig.~\ref{fig:D68percentile_MH}). However, this observed fraction constitutes a lower limit, as the most centrally concentrated populations in the Milky Way are inevitably most extincted, and hence most underrepresented in the Gaia XP sample.}
  \label{fig:sims_extent}
\end{figure*}

We start by asking how centrally concentrated the most metal-rich stars are in the simulated galaxies. This is illustrated in the left panel of Figure~\ref{fig:sims_extent}, which shows the Galactocentric radius, $R_{\mathrm{GC}}$ that encloses 68\% of stars in bins of increasing [M/H].
While for metallicities $\rm [M/H]\lesssim 0$ this 68\%-percentile radius is $\sim 5.5$~kpc, it drops in all cases quite dramatically to $<1$~kpc at high metallicities ($\rm [M/H] \gtrsim +0.2$). Both the metallicity at which the spatial extent drops so precipitously and the radius to which it drops, vary somewhat among the inspected systems: e.g., in half the cases the bulk of the highest metallicity stars is enclosed within 500~pc. This implies that we should indeed expect the most metal-rich stars in Milky-Way-like galaxies to be almost exclusively found at the very center. 
In these simulated galaxies, the metal rich stars of compact spatial extent form dynamically hot systems, with orbital circularities of $\eta\equiv L_z/L_{circ}\sim 0.4-0.6$ \citep{Naidu2020,Chandra2023}, in at least qualitative agreement with the observations.  

However, TNG50 offers no indication that the spatial distribution of the very- and the extremely-metal-rich stars differs, as seen in our Galaxy. In the TNG50 Milky Way analogs, stars over a much wider metallicity range are found predominately in a compact central core, typically within $\Delta[M/H]=0.4$ of the maximal metallicity reached. As the right panel of Figure~\ref{fig:sims_extent} shows, these metal-rich \things in the simulated systems also contain a much larger fraction of stars, typically 5-10\%, as opposed to $\le 0.1\%$ seen in our Milky Way sample with XP metallicities. For each simulation, we considered narrow mono-metallicity bins and calculated the radius that enclosed 68\% of all such stars (whithin 15~kpc). Then we asked what the lowest metallicity bin is for which this radius is $\le 1.5$~kpc, and calculated the fraction of the stellar metallicity distribution (MDF) that is higher than value. We take this to be the fraction of the stellar MDF that is ``predominately confined to a metal-rich core" in the simulations. The gray histogram in the right panel of Figure~\ref{fig:sims_extent} shows the distribution of this MDF fraction across all simulations considered.  This fraction varies across simulated analogs by an order of magnitude, from 2.5\% to 25\%. However, in none of the simulations is this fraction as minute as in the Gaia XP data. Considering the \emph{fraction} of the most metal-rich stars in the data-simulation comparison (rather than the absolute \MH ) has the advantage of minimizing the impact of any systematic differences in the metallicity scales.

\begin{figure}
  \includegraphics[width=1\columnwidth]{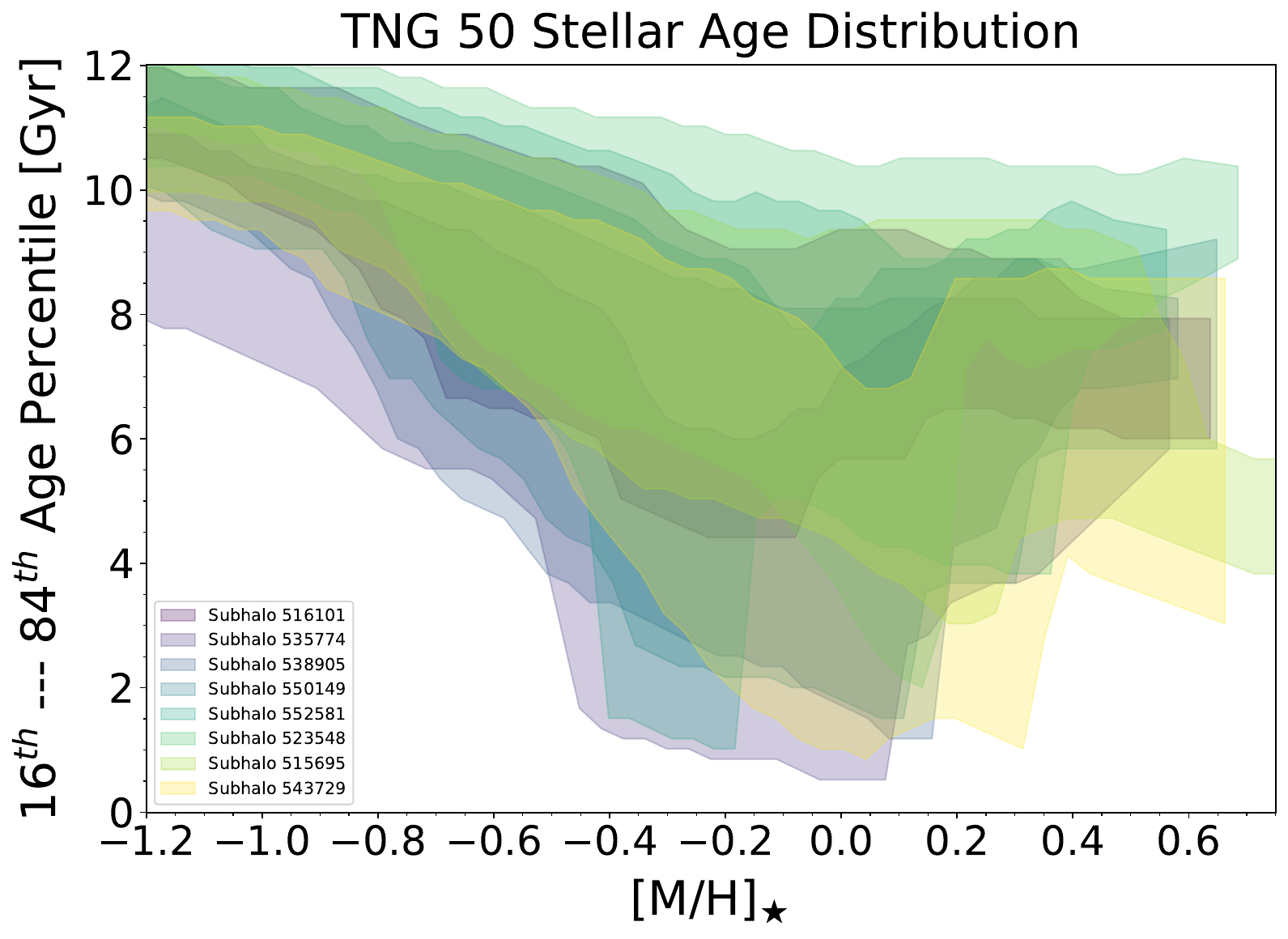}
  \caption{Age distribution of different mono-abundance stellar populations for a handful of Milky Way analogs selected from the TNG50 simulation \citep{Pillepich2023}. Specifically, we show the $16^{th}$ and $84^{th}$ percentiles of the stellar ages as a function of \MH. According to TNG50, the stars in the metal rich knot have a wide range of ages, $2\mathrm{Gyr}\lesssim \tau\lesssim 10\mathrm{Gyr}$, just as indicated by the data for the Milky Way (Fig.~\ref{fig:RGB_in_CMD} and  \citet{Bensby2017}).}
  \label{fig:sims_ages}
\end{figure}

We can also ask \emph{when} the most metal-rich stars in these simulations formed. This is illustrated in Figure~\ref{fig:sims_ages}, which shows the $16^{th}-84^{th}$ percentile age range of the stellar populations at different metallicities. This figure shows that in all systems the EMR stars formed over an extended period that ranges from 4 to 11 Gyrs among the subhalos. This matches the preliminary analysis of the ages of the Galactic EMR \thing (\S\ref{sec:ages}). And, it implies that \emph{several} episodes, typically near the peak of the overall star-formation rate for such galaxies (at $z\sim 1$) contributed to the central VMR and EMR population.
 TNG50 Milky Way analogs (e.g. Figure 12) in \citealt{Pillepich2023}) seem to make it plausible that several EMR-formation episodes occurred in the life of a single galaxy.

\section{Summary and Discussion}

We have examined the spatial distribution of the most metal-rich stars in the Milky Way by making stellar mono-abundance maps of giants with metallicities, $\MHXP$ from Gaia XP spectra. We have found that the extremely metal-rich (EMR, $\MHXP\gtrsim +0.5$) stars in the inner Galaxy -- but only those -- are tightly and predominately confined to a spheroidal stellar \thing  at $R_{\mathrm{GC}} \le 1.5$~kpc; only a small fraction of them seems to be at larger radii. 
Stars just slightly less metal rich (VMR stars with $0.2\le\MHXP\le 0.4$) have a distinctly different spatial, or on-sky, distribution. A fraction of them still have the same centrally concentrated distribution as this \thing of EMR stars, 
forming an overall metal-rich compact \thing with stars $0.2\lesssim \MHXP\lesssim 0.6$, with $p(\,\MHXP)$ peaking at $+0.35$. This \thing is presumably much the same as the component recently pointed out by \citet{Horta2024}. Our results here simply provide clearer evidence for the spatial distribution of these central VMR stars, discussed previously \citep{Ness2016a, Queiroz21, Johnson2022}.
However, the majority of these VMR stars form a flattened configuration of several kpc extent. 

We also find that the stars of the EMR \thing form a dynamically hot system with only modest net rotation and presumably radial orbits. Moreover, the very metal-rich stars in that \thing appear to have a wide range of ages, from 2-10~Gyrs. Higher parallax precision or spectroscopic age indicators would be needed to make a more definitive determination of the age distribution. 

A first comparison in Figure~\ref{fig:sims_extent} with Milky Way analogs in the TNG50 simulation shows that some, but not all, of these properties should be expected. In simulated Milky Way-like systems, the most metal-rich stars are almost exclusively found at the very center, in the inner 1-2 kpc.
These metal-rich \things are dynamically hot systems with typical orbit circularities of $\eta\sim 0.4-0.6$ (as defined in \citealt{Chandra2023}). And these stars have a range of ages (a few Gyrs, Figure~\ref{fig:sims_ages}), as they formed over multiple star-formation episodes. These aspects realized by the TNG50 simulation are in accordance with what we see in the Milky Way.

However, there is one important feature in which the TNG50 simulated Milky Way analogs differ markedly from what we have uncovered in the Milky Way with this paper. 
In TNG50, metal-rich populations confined to the central \thing ($R_{\mathrm{GC}}\lesssim 1.5$~kpc for 68\% of them) invariably encompass a significant fraction of the total stellar mass (Figure~\ref{fig:sims_extent}, right), typically 5-10\%, and seemingly never $<2\%$. 
This is in contrast to the only $\sim 0.1$\% of most metal rich stellar populations (i.e. the EMR stars) that are confined to such a \thing in our Galaxy\footnote{More explicitly:  the only mono-abundance sub-populations in the Gaia XP sample that are predominately confined to the $\lesssim 1.5$~kpc \thing are the 0.1\% most metal-rich sample members.}.
For comparison between the simulation and the Milky Way, we use here the \emph{percentiles} of stars metal rich enough so that their mono-abundance population is confined in a central \thingns, as this measure may be less sensitive to systematic differences in the \MH~ scale. 

Just to reiterate: in the subset of TNG50 galaxies that we deem to be reasonable analogs of our Galaxy, the 5-10\% most metal-rich populations are confined to a $\lesssim 1.5$~kpc \thing, while in the Milky Way the 5-10\% most metal rich populations (with $\MHXP$ of $\ge +0.25$ and +0.2, respectively) form mostly a much more extended, flattened configuration.  As a consequence, none of the simulated galaxies shows a dramatic change of structure within the few percent of most metal-rich stars, as the Milky Way does between its VMR and EMR stars. It seems unlikely that any differences in the bulge-to-disk ratios between the simulated galaxies \citep[e.g.][]{Zana2022} and the Milky Way \citep[they are at best modest][]{Zana2022} can explain this discrepancy.  

A similar picture emerges if we look at the actual metallicity ranges: in TNG50, the populations within $\Delta\MH \sim 0.4$~ dex of the maximum metallicity are typically confined to a central $\lesssim 1.5$~kpc \thingns (see Figure~\ref{fig:sims_extent}); 
 this is a much larger range than seen in the Milky Way (Figure~\ref{fig:D68percentile_MH}).

It must be noted that the $\sim$0.1\% value quoted above and based on the Gaia XP data (Figure~\ref{fig:sims_extent}) presumably constitutes a lower limit on the true fraction in the Milky Way. Clearly, the most centrally confined populations in the Milky Way will suffer a higher level of line-of-sight extinction that precludes their membership in the Gaia XP sample. But even if the Milky Way's EMR population was 5$\times$ underrepresented in the Gaia XP sample compared to the VMR populations, the Milky Way would remain exceptional (vis-a-vis the TNG50 simulations) in the tiny fraction of most metal-rich stars confined to the \thingns.

In summary, the discovery of an EMR \thingns, almost exclusively confined to near the Galactic center, is the most striking observational result presented here. However, the fact that the slightly less extreme VMR stars have such an extended distribution in our Galaxy is the most surprising, or at least most discrepant aspect, in comparison with the TNG50 simulated galaxies.


This state of affairs suggests a variety of follow-up observations and considerations.

Gaia data are severely limited by dust extinction, and new observations by SDSS-V (in conjunction with SDSS-IV data) will allow us to probe the very center of this \thing within a few degrees of the Galactic Center. The SDSS-V data will also allow us to link these data to existing studies with near-IR spectroscopy in the $\le 50$~pc of the Galactic center \citep[eg.][]{Feldmeier2017,Feldmeier2022}. 

Orbit and CMD analyses are currently limited by distance precision and dust extinction, and better spectrophotometric distances will help here. Spectroscopic giant ages (e.g. from their [C/N] ratio) would be important for better and independent constraints on the ages. But these indicators are currently not calibrated for stars of supersolar abundances. Finally, detailed abundances from SDSS's APOGEE spectra may elucidate the circumstances under which such high metallicities were reached.  The reason why the distribution of EMR stars in the Milky Way is so much more compact than that of slightly less metal-rich stars could be sorted out with specially designed high-resolution formation simulations.

Here, we have only considered one set of formation simulations. It would be interesting to use these face-value results to see whether they shed light on different star formation modes or of different numerical implementations of star formation and enrichment, along the lines of much existing and ongoing work on the Milky Way's particular formation history \citep[eg.][]{Amarante2022,Debattista2023}.

The existence of the Milky Way's EMR \thing offers a laboratory for studying the most metal-rich end of star formation in galaxies like ours. Is there a sharp cutoff in the stellar metallicity distribution and, if so, what sets its value? Why were such high metallicities reached across a wide range of epochs? Why can these EMR values only be reached in the inner 1~kpc? Why does the Milky Way have so many VMR stars far beyond 1.5~kpc, while typically the most metal rich 10\% of stars in TNG50 Milky Way analogs are confined to the central 1.5~kpc?

\section*{Acknowledgments}

This work has made use of data from the European Space Agency (ESA) mission
{\it Gaia} (\url{https://www.cosmos.esa.int/gaia}), processed by the {\it Gaia}
Data Processing and Analysis Consortium (DPAC,
\url{https://www.cosmos.esa.int/web/gaia/dpac/consortium}). Funding for the DPAC
has been provided by national institutions, in particular the institutions
participating in the {\it Gaia} Multilateral Agreement.

Funding for the Sloan Digital Sky Survey V (and its earlier phases such as SDSS-IV) has been provided by the Alfred P. Sloan Foundation, the Heising-Simons Foundation, the National Science Foundation, and the Participating Institutions. SDSS acknowledges support and resources from the Center for High-Performance Computing at the University of Utah. The SDSS web site is \url{www.sdss.org}. 

SDSS is managed by the Astrophysical Research Consortium for the Participating Institutions of the SDSS Collaboration, including the Carnegie Institution for Science, Chilean National Time Allocation Committee (CNTAC) ratified researchers, the Gotham Participation Group, Harvard University, Heidelberg University, 
The Johns Hopkins University, L’Ecole polytechnique f\'{e}d\'{e}rale de Lausanne (EPFL), Leibniz Institut for Astrophysics Potsdam (AIP), Max Planck Institute for Astronomy (MPIA Heidelberg), Max Planck Institut for Extraterrestrial Physics (MPE), Nanjing University, National Astronomical Observatories of China (NAOC),
New Mexico State University, The Ohio State University, Pennsylvania State University, 
Smithsonian Astrophysical Observatory, Space Telescope Science Institute (STScI), the Stellar Astrophysics Participation Group, Universidad Nacional Aut\'{o}noma de M\'{e}xico, University of Arizona, University of Colorado Boulder, University of Illinois at Urbana-Champaign,
University of Toronto, University of Utah, University of Virginia, Yale University, and Yunnan University.
NF acknowledges the support of the Natural Sciences and Engineering Research Council of Canada (NSERC), [funding reference number 568580] through a CITA postdoctoral fellowship, and acknowledges partial support from an Arts \& Sciences Postdoctoral Fellowship at the University of Toronto.







\bibliography{RCriches}

\begin{thebibliography}{}
\expandafter\ifx\csname natexlab\endcsname\relax\def\natexlab#1{#1}\fi
\providecommand{\url}[1]{\href{#1}{#1}}
\providecommand{\dodoi}[1]{doi:~\href{http://doi.org/#1}{\nolinkurl{#1}}}
\providecommand{\doeprint}[1]{\href{http://ascl.net/#1}{\nolinkurl{http://ascl.net/#1}}}
\providecommand{\doarXiv}[1]{\href{https://arxiv.org/abs/#1}{\nolinkurl{https://arxiv.org/abs/#1}}}

\bibitem[{{Almeida} {et~al.}(2023){Almeida}, {Anderson}, {Argudo-Fern{\'a}ndez}, {Badenes}, {Barger}, {Barrera-Ballesteros}, {Bender}, {Benitez}, {Besser}, {Bird}, {Bizyaev}, {Blanton}, {Bochanski}, {Bovy}, {Brandt}, {Brownstein}, {Buchner}, {Bulbul}, {Burchett}, {Cano D{\'\i}az}, {Carlberg}, {Casey}, {Chandra}, {Cherinka}, {Chiappini}, {Coker}, {Comparat}, {Conroy}, {Contardo}, {Cortes}, {Covey}, {Crane}, {Cunha}, {Dabbieri}, {Davidson}, {Davis}, {de Andrade Queiroz}, {De Lee}, {M{\'e}ndez Delgado}, {Demasi}, {Di Mille}, {Donor}, {Dow}, {Dwelly}, {Eracleous}, {Eriksen}, {Fan}, {Farr}, {Frederick}, {Fries}, {Frinchaboy}, {G{\"a}nsicke}, {Ge}, {Gonz{\'a}lez {\'A}vila}, {Grabowski}, {Grier}, {Guiglion}, {Gupta}, {Hall}, {Hawkins}, {Hayes}, {Hermes}, {Hern{\'a}ndez-Garc{\'\i}a}, {Hogg}, {Holtzman}, {Ibarra-Medel}, {Ji}, {Jofre}, {Johnson}, {Jones}, {Kinemuchi}, {Kluge}, {Koekemoer}, {Kollmeier}, {Kounkel}, {Krishnarao}, {Krumpe}, {Lacerna}, {Lago}, {Laporte}, {Liu}, {Liu}, {Liu}, {Lopes}, {Macktoobian},
  {Majewski}, {Malanushenko}, {Maoz}, {Masseron}, {Masters}, {Matijevic}, {McBride}, {Medan}, {Merloni}, {Morrison}, {Myers}, {M{\'e}sz{\'a}ros}, {Negrete}, {Nidever}, {Nitschelm}, {Oravetz}, {Oravetz}, {Pan}, {Peng}, {Pinsonneault}, {Pogge}, {Qiu}, {Ramirez}, {Rix}, {Fern{\'a}ndez Rosso}, {Runnoe}, {Salvato}, {Sanchez}, {Santana}, {Saydjari}, {Sayres}, {Schlaufman}, {Schneider}, {Schwope}, {Serna}, {Shen}, {Sobeck}, {Song}, {Souto}, {Spoo}, {Stassun}, {Steinmetz}, {Straumit}, {Stringfellow}, {S{\'a}nchez-Gallego}, {Taghizadeh-Popp}, {Tayar}, {Thakar}, {Tissera}, {Tkachenko}, {Hernandez Toledo}, {Trakhtenbrot}, {Fern{\'a}ndez-Trincado}, {Troup}, {Trump}, {Tuttle}, {Ulloa}, {Vazquez-Mata}, {Vera Alfaro}, {Villanova}, {Wachter}, {Weijmans}, {Wheeler}, {Wilson}, {Wojno}, {Wolf}, {Xue}, {Ybarra}, {Zari}, \& {Zasowski}}]{Almeida2023}
{Almeida}, A., {Anderson}, S.~F., {Argudo-Fern{\'a}ndez}, M., {et~al.} 2023, \apjs, 267, 44, \dodoi{10.3847/1538-4365/acda98}

\bibitem[{{Amarante} {et~al.}(2022){Amarante}, {Debattista}, {Beraldo e Silva}, {Laporte}, \& {Deg}}]{Amarante2022}
{Amarante}, J. A.~S., {Debattista}, V.~P., {Beraldo e Silva}, L., {Laporte}, C. F.~P., \& {Deg}, N. 2022, \apj, 937, 12, \dodoi{10.3847/1538-4357/ac8b0d}

\bibitem[{{Andrae} {et~al.}(2023{\natexlab{a}}){Andrae}, {Rix}, \& {Chandra}}]{Andrae23b}
{Andrae}, R., {Rix}, H.-W., \& {Chandra}, V. 2023{\natexlab{a}}, \apjs, 267, 8, \dodoi{10.3847/1538-4365/acd53e}

\bibitem[{{Andrae} {et~al.}(2023{\natexlab{b}}){Andrae}, {Rix}, \& {Chandra}}]{Andrae2023}
---. 2023{\natexlab{b}}, \apjs, 267, 8, \dodoi{10.3847/1538-4365/acd53e}

\bibitem[{{Andrae} {et~al.}(2023{\natexlab{c}}){Andrae}, {Fouesneau}, {Sordo}, {Bailer-Jones}, {Dharmawardena}, {Rybizki}, {De Angeli}, {Lindstr{\o}m}, {Marshall}, {Drimmel}, {Korn}, {Soubiran}, {Brouillet}, {Casamiquela}, {Rix}, {Abreu Aramburu}, {{\'A}lvarez}, {Bakker}, {Bellas-Velidis}, {Bijaoui}, {Brugaletta}, {Burlacu}, {Carballo}, {Chaoul}, {Chiavassa}, {Contursi}, {Cooper}, {Creevey}, {Dafonte}, {Dapergolas}, {de Laverny}, {Delchambre}, {Demouchy}, {Edvardsson}, {Fr{\'e}mat}, {Garabato}, {Garc{\'\i}a-Lario}, {Garc{\'\i}a-Torres}, {Gavel}, {Gomez}, {Gonz{\'a}lez-Santamar{\'\i}a}, {Hatzidimitriou}, {Heiter}, {Jean-Antoine Piccolo}, {Kontizas}, {Kordopatis}, {Lanzafame}, {Lebreton}, {Licata}, {Livanou}, {Lobel}, {Lorca}, {Magdaleno Romeo}, {Manteiga}, {Marocco}, {Mary}, {Nicolas}, {Ordenovic}, {Pailler}, {Palicio}, {Pallas-Quintela}, {Panem}, {Pichon}, {Poggio}, {Recio-Blanco}, {Riclet}, {Robin}, {Santove{\~n}a}, {Sarro}, {Schultheis}, {Segol}, {Silvelo}, {Slezak}, {Smart}, {S{\"u}veges}, {Th{\'e}venin},
  {Torralba Elipe}, {Ulla}, {Utrilla}, {Vallenari}, {van Dillen}, {Zhao}, \& {Zorec}}]{Andrae23a}
{Andrae}, R., {Fouesneau}, M., {Sordo}, R., {et~al.} 2023{\natexlab{c}}, \aap, 674, A27, \dodoi{10.1051/0004-6361/202243462}

\bibitem[{{Arentsen} {et~al.}(2020){Arentsen}, {Starkenburg}, {Martin}, {Hill}, {Ibata}, {Kunder}, {Schultheis}, {Venn}, {Zucker}, {Aguado}, {Carlberg}, {Gonz{\'a}lez Hern{\'a}ndez}, {Lardo}, {Longeard}, {Malhan}, {Navarro}, {S{\'a}nchez-Janssen}, {Sestito}, {Thomas}, {Youakim}, {Lewis}, {Simpson}, \& {Wan}}]{Arentsen2020}
{Arentsen}, A., {Starkenburg}, E., {Martin}, N.~F., {et~al.} 2020, \mnras, 491, L11, \dodoi{10.1093/mnrasl/slz156}

\bibitem[{{Belokurov} {et~al.}(2018){Belokurov}, {Erkal}, {Evans}, {Koposov}, \& {Deason}}]{Belokurov_disk_halo}
{Belokurov}, V., {Erkal}, D., {Evans}, N.~W., {Koposov}, S.~E., \& {Deason}, A.~J. 2018, \mnras, 478, 611, \dodoi{10.1093/mnras/sty982}

\bibitem[{{Belokurov} \& {Kravtsov}(2022)}]{BelokurovKravtsov22}
{Belokurov}, V., \& {Kravtsov}, A. 2022, \mnras, 514, 689, \dodoi{10.1093/mnras/stac1267}

\bibitem[{{Belokurov} {et~al.}(2020){Belokurov}, {Sanders}, {Fattahi}, {Smith}, {Deason}, {Evans}, \& {Grand}}]{Belokurov_splash}
{Belokurov}, V., {Sanders}, J.~L., {Fattahi}, A., {et~al.} 2020, \mnras, 494, 3880, \dodoi{10.1093/mnras/staa876}

\bibitem[{{Bensby} {et~al.}(2017){Bensby}, {Feltzing}, {Gould}, {Yee}, {Johnson}, {Asplund}, {Mel{\'e}ndez}, {Lucatello}, {Howes}, {McWilliam}, {Udalski}, {Szyma{\'n}ski}, {Soszy{\'n}ski}, {Poleski}, {Wyrzykowski}, {Ulaczyk}, {Koz{\l}owski}, {Pietrukowicz}, {Skowron}, {Mr{\'o}z}, {Pawlak}, {Abe}, {Asakura}, {Bhattacharya}, {Bond}, {Bennett}, {Hirao}, {Nagakane}, {Koshimoto}, {Sumi}, {Suzuki}, \& {Tristram}}]{Bensby2017}
{Bensby}, T., {Feltzing}, S., {Gould}, A., {et~al.} 2017, \aap, 605, A89, \dodoi{10.1051/0004-6361/201730560}

\bibitem[{{Bernard} {et~al.}(2018){Bernard}, {Schultheis}, {Di Matteo}, {Hill}, {Haywood}, \& {Calamida}}]{Bernard2018}
{Bernard}, E.~J., {Schultheis}, M., {Di Matteo}, P., {et~al.} 2018, \mnras, 477, 3507, \dodoi{10.1093/mnras/sty902}

\bibitem[{{Binney}(2010)}]{Binney2010}
{Binney}, J. 2010, \mnras, 401, 2318, \dodoi{10.1111/j.1365-2966.2009.15845.x}

\bibitem[{{Binney} \& {Tremaine}(2008)}]{BT2008}
{Binney}, J., \& {Tremaine}, S. 2008, {Galactic Dynamics: Second Edition}

\bibitem[{{Bland-Hawthorn} \& {Gerhard}(2016)}]{BlandHawthorn2016}
{Bland-Hawthorn}, J., \& {Gerhard}, O. 2016, \araa, 54, 529, \dodoi{10.1146/annurev-astro-081915-023441}

\bibitem[{{Boecker} {et~al.}(2023){Boecker}, {Neumayer}, {Pillepich}, {Frankel}, {Ramesh}, {Leaman}, \& {Hernquist}}]{Boecker2023}
{Boecker}, A., {Neumayer}, N., {Pillepich}, A., {et~al.} 2023, \mnras, 519, 5202, \dodoi{10.1093/mnras/stac3759}

\bibitem[{{Bressan} {et~al.}(2012){Bressan}, {Marigo}, {Girardi}, {Salasnich}, {Dal Cero}, {Rubele}, \& {Nanni}}]{Bressan2012}
{Bressan}, A., {Marigo}, P., {Girardi}, L., {et~al.} 2012, \mnras, 427, 127, \dodoi{10.1111/j.1365-2966.2012.21948.x}

\bibitem[{{Chandra} {et~al.}(2023){Chandra}, {Semenov}, {Rix}, {Conroy}, {Bonaca}, {Naidu}, {Andrae}, {Li}, \& {Hernquist}}]{Chandra2023}
{Chandra}, V., {Semenov}, V.~A., {Rix}, H.-W., {et~al.} 2023, arXiv e-prints, arXiv:2310.13050, \dodoi{10.48550/arXiv.2310.13050}

\bibitem[{{Chiti} {et~al.}(2021){Chiti}, {Mardini}, {Frebel}, \& {Daniel}}]{Chiti2021}
{Chiti}, A., {Mardini}, M.~K., {Frebel}, A., \& {Daniel}, T. 2021, ApJL, 911, L23, \dodoi{10.3847/2041-8213/abd629}

\bibitem[{{Cirasuolo} {et~al.}(2020){Cirasuolo}, {Fairley}, {Rees}, {Gonzalez}, {Taylor}, {Maiolino}, {Afonso}, {Evans}, {Flores}, {Lilly}, {Oliva}, {Paltani}, {Vanzi}, {Abreu}, {Accardo}, {Adams}, {{\'A}lvarez M{\'e}ndez}, {Amans}, {Amarantidis}, {Atek}, {Atkinson}, {Banerji}, {Barrett}, {Barrientos}, {Bauer}, {Beard}, {B{\'e}chet}, {Belfiore}, {Bellazzini}, {Benoist}, {Best}, {Biazzo}, {Black}, {Boettger}, {Bonifacio}, {Bowler}, {Bragaglia}, {Brierley}, {Brinchmann}, {Brinkmann}, {Buat}, {Buitrago}, {Burgarella}, {Burningham}, {Buscher}, {Cabral}, {Caffau}, {Cardoso}, {Carnall}, {Carollo}, {Castillo}, {Castignani}, {Catelan}, {Cicone}, {Cimatti}, {Cioni}, {Clementini}, {Cochrane}, {Coelho}, {Colling}, {Contini}, {Contreras}, {Conzelmann}, {Cresci}, {Cropper}, {Cucciati}, {Cullen}, {Cumani}, {Curti}, {Da Silva}, {Daddi}, {Dalessandro}, {Dalessio}, {Dauvin}, {Davidson}, {de Laverny}, {Delplancke-Str{\"o}bele}, {De Lucia}, {Del Vecchio}, {Dessauges-Zavadsky}, {Di Matteo}, {Dole}, {Drass}, {Dunlop},
  {D{\"u}nner}, {Eales}, {Ellis}, {Enriques}, {Fasola}, {Ferguson}, {Ferruzzi}, {Fisher}, {Flores}, {Fontana}, {Forchi}, {Francois}, {Franzetti}, {Gargiulo}, {Garilli}, {Gaudemard}, {Gieles}, {Gilmore}, {Ginolfi}, {Gomes}, {Guinouard}, {Gutierrez}, {Haigron}, {Hammer}, {Hammersley}, {Haniff}, {Harrison}, {Haywood}, {Hill}, {Hubin}, {Humphrey}, {Ibata}, {Infante}, {Ives}, {Ivison}, {Iwert}, {Jablonka}, {Jakob}, {Jarvis}, {King}, {Kneib}, {Laporte}, {Lawrence}, {Lee}, {Li Causi}, {Lorenzoni}, {Lucatello}, {Luco}, {Macleod}, {Magliocchetti}, {Magrini}, {Mainieri}, {Maire}, {Mannucci}, {Martin}, {Matute}, {Maurogordato}, {McGee}, {Mcleod}, {McLure}, {McMahon}, {Melse}, {Messias}, {Mucciarelli}, {Nisini}, {Nix}, {Norberg}, {Oesch}, {Oliveira}, {Origlia}, {Padilla}, {Palsa}, {Pancino}, {Papaderos}, {Pappalardo}, {Parry}, {Pasquini}, {Peacock}, {Pedichini}, {Pello}, {Peng}, {Pentericci}, {Pfuhl}, {Piazzesi}, {Popovic}, {Pozzetti}, {Puech}, {Puzia}, {Raichoor}, {Randich}, {Recio-Blanco}, {Reis}, {Reix}, {Renzini},
  {Rodrigues}, {Rojas}, {Rojas-Arriagada}, {Rota}, {Royer}, {Sacco}, {Sanchez-Janssen}, {Sanna}, {Santos}, {Sarzi}, {Schaerer}, {Schiavon}, {Schnell}, {Schultheis}, {Scodeggio}, {Serjeant}, {Shen}, {Simmonds}, {Smoker}, {Sobral}, {Sordet}, {Sp{\'e}rone}, {Strachan}, {Sun}, {Swinbank}, {Tait}, {Tereno}, {Tojeiro}, {Torres}, {Tosi}, {Tozzi}, {Tresiter}, {Valenti}, {Valenzuela Navarro}, {Vanzella}, {Vergani}, {Verhamme}, {Vernet}, {Vignali}, {Vinther}, {Von Dran}, {Waring}, {Watson}, {Wild}, {Willesme}, {Woodward}, {Wuyts}, {Yang}, {Zamorani}, {Zoccali}, {Bluck}, \& {Trussler}}]{Cirasuolo2020}
{Cirasuolo}, M., {Fairley}, A., {Rees}, P., {et~al.} 2020, The Messenger, 180, 10, \dodoi{10.18727/0722-6691/5195}

\bibitem[{{De Angeli} {et~al.}(2023){De Angeli}, {Weiler}, {Montegriffo}, {Evans}, {Riello}, {Andrae}, {Carrasco}, {Busso}, {Burgess}, {Cacciari}, {Davidson}, {Harrison}, {Hodgkin}, {Jordi}, {Osborne}, {Pancino}, {Altavilla}, {Barstow}, {Bailer-Jones}, {Bellazzini}, {Brown}, {Castellani}, {Cowell}, {Delchambre}, {De Luise}, {Diener}, {Fabricius}, {Fouesneau}, {Fr{\'e}mat}, {Gilmore}, {Giuffrida}, {Hambly}, {Hidalgo}, {Holland}, {Kostrzewa-Rutkowska}, {van Leeuwen}, {Lobel}, {Marinoni}, {Miller}, {Pagani}, {Palaversa}, {Piersimoni}, {Pulone}, {Ragaini}, {Rainer}, {Richards}, {Rixon}, {Ruz-Mieres}, {Sanna}, {Sarro}, {Rowell}, {Sordo}, {Walton}, \& {Yoldas}}]{DeAngeli2023}
{De Angeli}, F., {Weiler}, M., {Montegriffo}, P., {et~al.} 2023, A\&A, 674, A2, \dodoi{10.1051/0004-6361/202243680}

\bibitem[{{de Jong} {et~al.}(2019){de Jong}, {Agertz}, {Berbel}, {Aird}, {Alexander}, {Amarsi}, {Anders}, {Andrae}, {Ansarinejad}, {Ansorge}, {Antilogus}, {Anwand-Heerwart}, {Arentsen}, {Arnadottir}, {Asplund}, {Auger}, {Azais}, {Baade}, {Baker}, {Baker}, {Balbinot}, {Baldry}, {Banerji}, {Barden}, {Barklem}, {Barth{\'e}l{\'e}my-Mazot}, {Battistini}, {Bauer}, {Bell}, {Bellido-Tirado}, {Bellstedt}, {Belokurov}, {Bensby}, {Bergemann}, {Bestenlehner}, {Bielby}, {Bilicki}, {Blake}, {Bland-Hawthorn}, {Boeche}, {Boland}, {Boller}, {Bongard}, {Bongiorno}, {Bonifacio}, {Boudon}, {Brooks}, {Brown}, {Brown}, {Br{\"u}ggen}, {Brynnel}, {Brzeski}, {Buchert}, {Buschkamp}, {Caffau}, {Caillier}, {Carrick}, {Casagrande}, {Case}, {Casey}, {Cesarini}, {Cescutti}, {Chapuis}, {Chiappini}, {Childress}, {Christlieb}, {Church}, {Cioni}, {Cluver}, {Colless}, {Collett}, {Comparat}, {Cooper}, {Couch}, {Courbin}, {Croom}, {Croton}, {Daguis{\'e}}, {Dalton}, {Davies}, {Davis}, {de Laverny}, {Deason}, {Dionies}, {Disseau}, {Doel},
  {D{\"o}scher}, {Driver}, {Dwelly}, {Eckert}, {Edge}, {Edvardsson}, {Youssoufi}, {Elhaddad}, {Enke}, {Erfanianfar}, {Farrell}, {Fechner}, {Feiz}, {Feltzing}, {Ferreras}, {Feuerstein}, {Feuillet}, {Finoguenov}, {Ford}, {Fotopoulou}, {Fouesneau}, {Frenk}, {Frey}, {Gaessler}, {Geier}, {Gentile Fusillo}, {Gerhard}, {Giannantonio}, {Giannone}, {Gibson}, {Gillingham}, {Gonz{\'a}lez-Fern{\'a}ndez}, {Gonzalez-Solares}, {Gottloeber}, {Gould}, {Grebel}, {Gueguen}, {Guiglion}, {Haehnelt}, {Hahn}, {Hansen}, {Hartman}, {Hauptner}, {Hawkins}, {Haynes}, {Haynes}, {Heiter}, {Helmi}, {Aguayo}, {Hewett}, {Hinton}, {Hobbs}, {Hoenig}, {Hofman}, {Hook}, {Hopgood}, {Hopkins}, {Hourihane}, {Howes}, {Howlett}, {Huet}, {Irwin}, {Iwert}, {Jablonka}, {Jahn}, {Jahnke}, {Jarno}, {Jin}, {Jofre}, {Johl}, {Jones}, {J{\"o}nsson}, {Jordan}, {Karovicova}, {Khalatyan}, {Kelz}, {Kennicutt}, {King}, {Kitaura}, {Klar}, {Klauser}, {Kneib}, {Koch}, {Koposov}, {Kordopatis}, {Korn}, {Kosmalski}, {Kotak}, {Kovalev}, {Kreckel}, {Kripak}, {Krumpe},
  {Kuijken}, {Kunder}, {Kushniruk}, {Lam}, {Lamer}, {Laurent}, {Lawrence}, {Lehmitz}, {Lemasle}, {Lewis}, {Li}, {Lidman}, {Lind}, {Liske}, {Lizon}, {Loveday}, {Ludwig}, {McDermid}, {Maguire}, {Mainieri}, {Mali}, {Mandel}, {Mandel}, {Mannering}, {Martell}, {Martinez Delgado}, {Matijevic}, {McGregor}, {McMahon}, {McMillan}, {Mena}, {Merloni}, {Meyer}, {Michel}, {Micheva}, {Migniau}, {Minchev}, {Monari}, {Muller}, {Murphy}, {Muthukrishna}, {Nandra}, {Navarro}, {Ness}, {Nichani}, {Nichol}, {Nicklas}, {Niederhofer}, {Norberg}, {Obreschkow}, {Oliver}, {Owers}, {Pai}, {Pankratow}, {Parkinson}, {Paschke}, {Paterson}, {Pecontal}, {Parry}, {Phillips}, {Pillepich}, {Pinard}, {Pirard}, {Piskunov}, {Plank}, {Pl{\"u}schke}, {Pons}, {Popesso}, {Power}, {Pragt}, {Pramskiy}, {Pryer}, {Quattri}, {Queiroz}, {Quirrenbach}, {Rahurkar}, {Raichoor}, {Ramstedt}, {Rau}, {Recio-Blanco}, {Reiss}, {Renaud}, {Revaz}, {Rhode}, {Richard}, {Richter}, {Rix}, {Robotham}, {Roelfsema}, {Romaniello}, {Rosario}, {Rothmaier}, {Roukema}, {Ruchti},
  {Rupprecht}, {Rybizki}, {Ryde}, {Saar}, {Sadler}, {Sahl{\'e}n}, {Salvato}, {Sassolas}, {Saunders}, {Saviauk}, {Sbordone}, {Schmidt}, {Schnurr}, {Scholz}, {Schwope}, {Seifert}, {Shanks}, {Sheinis}, {Sivov}, {Sk{\'u}lad{\'o}ttir}, {Smartt}, {Smedley}, {Smith}, {Smith}, {Sorce}, {Spitler}, {Starkenburg}, {Steinmetz}, {Stilz}, {Storm}, {Sullivan}, {Sutherland}, {Swann}, {Tamone}, {Taylor}, {Teillon}, {Tempel}, {ter Horst}, {Thi}, {Tolstoy}, {Trager}, {Traven}, {Tremblay}, {Tresse}, {Valentini}, {van de Weygaert}, {van den Ancker}, {Veljanoski}, {Venkatesan}, {Wagner}, {Wagner}, {Walcher}, {Waller}, {Walton}, {Wang}, {Winkler}, {Wisotzki}, {Worley}, {Worseck}, {Xiang}, {Xu}, {Yong}, {Zhao}, {Zheng}, {Zscheyge}, \& {Zucker}}]{deJong2019}
{de Jong}, R.~S., {Agertz}, O., {Berbel}, A.~A., {et~al.} 2019, The Messenger, 175, 3, \dodoi{10.18727/0722-6691/5117}

\bibitem[{{De Silva} {et~al.}(2015){De Silva}, {Freeman}, {Bland-Hawthorn}, {Martell}, {de Boer}, {Asplund}, {Keller}, {Sharma}, {Zucker}, {Zwitter}, {Anguiano}, {Bacigalupo}, {Bayliss}, {Beavis}, {Bergemann}, {Campbell}, {Cannon}, {Carollo}, {Casagrande}, {Casey}, {Da Costa}, {D'Orazi}, {Dotter}, {Duong}, {Heger}, {Ireland}, {Kafle}, {Kos}, {Lattanzio}, {Lewis}, {Lin}, {Lind}, {Munari}, {Nataf}, {O'Toole}, {Parker}, {Reid}, {Schlesinger}, {Sheinis}, {Simpson}, {Stello}, {Ting}, {Traven}, {Watson}, {Wittenmyer}, {Yong}, \& {{\v{Z}}erjal}}]{DeSilva2015}
{De Silva}, G.~M., {Freeman}, K.~C., {Bland-Hawthorn}, J., {et~al.} 2015, \mnras, 449, 2604, \dodoi{10.1093/mnras/stv327}

\bibitem[{{Debattista} {et~al.}(2023){Debattista}, {Liddicott}, {Gonzalez}, {Beraldo e Silva}, {Amarante}, {Lazar}, {Zoccali}, {Valenti}, {Fisher}, {Khachaturyants}, {Nidever}, {Quinn}, {Du}, \& {Kassin}}]{Debattista2023}
{Debattista}, V.~P., {Liddicott}, D.~J., {Gonzalez}, O.~A., {et~al.} 2023, \apj, 946, 118, \dodoi{10.3847/1538-4357/acbb00}

\bibitem[{{Feldmeier-Krause}(2022)}]{Feldmeier2022}
{Feldmeier-Krause}, A. 2022, \mnras, 513, 5920, \dodoi{10.1093/mnras/stac1227}

\bibitem[{{Feldmeier-Krause} {et~al.}(2017){Feldmeier-Krause}, {Kerzendorf}, {Neumayer}, {Sch{\"o}del}, {Nogueras-Lara}, {Do}, {de Zeeuw}, \& {Kuntschner}}]{Feldmeier2017}
{Feldmeier-Krause}, A., {Kerzendorf}, W., {Neumayer}, N., {et~al.} 2017, \mnras, 464, 194, \dodoi{10.1093/mnras/stw2339}

\bibitem[{{Freeman} \& {Bland-Hawthorn}(2002)}]{Freeman2002}
{Freeman}, K., \& {Bland-Hawthorn}, J. 2002, ARA\&A, 40, 487, \dodoi{10.1146/annurev.astro.40.060401.093840}

\bibitem[{{Gaia Collaboration} {et~al.}(2016){Gaia Collaboration}, {Prusti}, {de Bruijne}, {Brown}, {Vallenari}, {Babusiaux}, {Bailer-Jones}, {Bastian}, {Biermann}, {Evans}, {Eyer}, {Jansen}, {Jordi}, {Klioner}, {Lammers}, {Lindegren}, {Luri}, {Mignard}, {Milligan}, {Panem}, {Poinsignon}, {Pourbaix}, {Randich}, {Sarri}, {Sartoretti}, {Siddiqui}, {Soubiran}, {Valette}, {van Leeuwen}, {Walton}, {Aerts}, {Arenou}, {Cropper}, {Drimmel}, {H{\o}g}, {Katz}, {Lattanzi}, {O'Mullane}, {Grebel}, {Holland}, {Huc}, {Passot}, {Bramante}, {Cacciari}, {Casta{\~n}eda}, {Chaoul}, {Cheek}, {De Angeli}, {Fabricius}, {Guerra}, {Hern{\'a}ndez}, {Jean-Antoine-Piccolo}, {Masana}, {Messineo}, {Mowlavi}, {Nienartowicz}, {Ord{\'o}{\~n}ez-Blanco}, {Panuzzo}, {Portell}, {Richards}, {Riello}, {Seabroke}, {Tanga}, {Th{\'e}venin}, {Torra}, {Els}, {Gracia-Abril}, {Comoretto}, {Garcia-Reinaldos}, {Lock}, {Mercier}, {Altmann}, {Andrae}, {Astraatmadja}, {Bellas-Velidis}, {Benson}, {Berthier}, {Blomme}, {Busso}, {Carry}, {Cellino}, {Clementini},
  {Cowell}, {Creevey}, {Cuypers}, {Davidson}, {De Ridder}, {de Torres}, {Delchambre}, {Dell'Oro}, {Ducourant}, {Fr{\'e}mat}, {Garc{\'\i}a-Torres}, {Gosset}, {Halbwachs}, {Hambly}, {Harrison}, {Hauser}, {Hestroffer}, {Hodgkin}, {Huckle}, {Hutton}, {Jasniewicz}, {Jordan}, {Kontizas}, {Korn}, {Lanzafame}, {Manteiga}, {Moitinho}, {Muinonen}, {Osinde}, {Pancino}, {Pauwels}, {Petit}, {Recio-Blanco}, {Robin}, {Sarro}, {Siopis}, {Smith}, {Smith}, {Sozzetti}, {Thuillot}, {van Reeven}, {Viala}, {Abbas}, {Abreu Aramburu}, {Accart}, {Aguado}, {Allan}, {Allasia}, {Altavilla}, {{\'A}lvarez}, {Alves}, {Anderson}, {Andrei}, {Anglada Varela}, {Antiche}, {Antoja}, {Ant{\'o}n}, {Arcay}, {Atzei}, {Ayache}, {Bach}, {Baker}, {Balaguer-N{\'u}{\~n}ez}, {Barache}, {Barata}, {Barbier}, {Barblan}, {Baroni}, {Barrado y Navascu{\'e}s}, {Barros}, {Barstow}, {Becciani}, {Bellazzini}, {Bellei}, {Bello Garc{\'\i}a}, {Belokurov}, {Bendjoya}, {Berihuete}, {Bianchi}, {Bienaym{\'e}}, {Billebaud}, {Blagorodnova}, {Blanco-Cuaresma}, {Boch},
  {Bombrun}, {Borrachero}, {Bouquillon}, {Bourda}, {Bouy}, {Bragaglia}, {Breddels}, {Brouillet}, {Br{\"u}semeister}, {Bucciarelli}, {Budnik}, {Burgess}, {Burgon}, {Burlacu}, {Busonero}, {Buzzi}, {Caffau}, {Cambras}, {Campbell}, {Cancelliere}, {Cantat-Gaudin}, {Carlucci}, {Carrasco}, {Castellani}, {Charlot}, {Charnas}, {Charvet}, {Chassat}, {Chiavassa}, {Clotet}, {Cocozza}, {Collins}, {Collins}, {Costigan}, {Crifo}, {Cross}, {Crosta}, {Crowley}, {Dafonte}, {Damerdji}, {Dapergolas}, {David}, {David}, {De Cat}, {de Felice}, {de Laverny}, {De Luise}, {De March}, {de Martino}, {de Souza}, {Debosscher}, {del Pozo}, {Delbo}, {Delgado}, {Delgado}, {di Marco}, {Di Matteo}, {Diakite}, {Distefano}, {Dolding}, {Dos Anjos}, {Drazinos}, {Dur{\'a}n}, {Dzigan}, {Ecale}, {Edvardsson}, {Enke}, {Erdmann}, {Escolar}, {Espina}, {Evans}, {Eynard Bontemps}, {Fabre}, {Fabrizio}, {Faigler}, {Falc{\~a}o}, {Farr{\`a}s Casas}, {Faye}, {Federici}, {Fedorets}, {Fern{\'a}ndez-Hern{\'a}ndez}, {Fernique}, {Fienga}, {Figueras}, {Filippi},
  {Findeisen}, {Fonti}, {Fouesneau}, {Fraile}, {Fraser}, {Fuchs}, {Furnell}, {Gai}, {Galleti}, {Galluccio}, {Garabato}, {Garc{\'\i}a-Sedano}, {Gar{\'e}}, {Garofalo}, {Garralda}, {Gavras}, {Gerssen}, {Geyer}, {Gilmore}, {Girona}, {Giuffrida}, {Gomes}, {Gonz{\'a}lez-Marcos}, {Gonz{\'a}lez-N{\'u}{\~n}ez}, {Gonz{\'a}lez-Vidal}, {Granvik}, {Guerrier}, {Guillout}, {Guiraud}, {G{\'u}rpide}, {Guti{\'e}rrez-S{\'a}nchez}, {Guy}, {Haigron}, {Hatzidimitriou}, {Haywood}, {Heiter}, {Helmi}, {Hobbs}, {Hofmann}, {Holl}, {Holland}, {Hunt}, {Hypki}, {Icardi}, {Irwin}, {Jevardat de Fombelle}, {Jofr{\'e}}, {Jonker}, {Jorissen}, {Julbe}, {Karampelas}, {Kochoska}, {Kohley}, {Kolenberg}, {Kontizas}, {Koposov}, {Kordopatis}, {Koubsky}, {Kowalczyk}, {Krone-Martins}, {Kudryashova}, {Kull}, {Bachchan}, {Lacoste-Seris}, {Lanza}, {Lavigne}, {Le Poncin-Lafitte}, {Lebreton}, {Lebzelter}, {Leccia}, {Leclerc}, {Lecoeur-Taibi}, {Lemaitre}, {Lenhardt}, {Leroux}, {Liao}, {Licata}, {Lindstr{\o}m}, {Lister}, {Livanou}, {Lobel}, {L{\"o}ffler},
  {L{\'o}pez}, {Lopez-Lozano}, {Lorenz}, {Loureiro}, {MacDonald}, {Magalh{\~a}es Fernandes}, {Managau}, {Mann}, {Mantelet}, {Marchal}, {Marchant}, {Marconi}, {Marie}, {Marinoni}, {Marrese}, {Marschalk{\'o}}, {Marshall}, {Mart{\'\i}n-Fleitas}, {Martino}, {Mary}, {Matijevi{\v{c}}}, {Mazeh}, {McMillan}, {Messina}, {Mestre}, {Michalik}, {Millar}, {Miranda}, {Molina}, {Molinaro}, {Molinaro}, {Moln{\'a}r}, {Moniez}, {Montegriffo}, {Monteiro}, {Mor}, {Mora}, {Morbidelli}, {Morel}, {Morgenthaler}, {Morley}, {Morris}, {Mulone}, {Muraveva}, {Musella}, {Narbonne}, {Nelemans}, {Nicastro}, {Noval}, {Ord{\'e}novic}, {Ordieres-Mer{\'e}}, {Osborne}, {Pagani}, {Pagano}, {Pailler}, {Palacin}, {Palaversa}, {Parsons}, {Paulsen}, {Pecoraro}, {Pedrosa}, {Pentik{\"a}inen}, {Pereira}, {Pichon}, {Piersimoni}, {Pineau}, {Plachy}, {Plum}, {Poujoulet}, {Pr{\v{s}}a}, {Pulone}, {Ragaini}, {Rago}, {Rambaux}, {Ramos-Lerate}, {Ranalli}, {Rauw}, {Read}, {Regibo}, {Renk}, {Reyl{\'e}}, {Ribeiro}, {Rimoldini}, {Ripepi}, {Riva}, {Rixon},
  {Roelens}, {Romero-G{\'o}mez}, {Rowell}, {Royer}, {Rudolph}, {Ruiz-Dern}, {Sadowski}, {Sagrist{\`a} Sell{\'e}s}, {Sahlmann}, {Salgado}, {Salguero}, {Sarasso}, {Savietto}, {Schnorhk}, {Schultheis}, {Sciacca}, {Segol}, {Segovia}, {Segransan}, {Serpell}, {Shih}, {Smareglia}, {Smart}, {Smith}, {Solano}, {Solitro}, {Sordo}, {Soria Nieto}, {Souchay}, {Spagna}, {Spoto}, {Stampa}, {Steele}, {Steidelm{\"u}ller}, {Stephenson}, {Stoev}, {Suess}, {S{\"u}veges}, {Surdej}, {Szabados}, {Szegedi-Elek}, {Tapiador}, {Taris}, {Tauran}, {Taylor}, {Teixeira}, {Terrett}, {Tingley}, {Trager}, {Turon}, {Ulla}, {Utrilla}, {Valentini}, {van Elteren}, {Van Hemelryck}, {van Leeuwen}, {Varadi}, {Vecchiato}, {Veljanoski}, {Via}, {Vicente}, {Vogt}, {Voss}, {Votruba}, {Voutsinas}, {Walmsley}, {Weiler}, {Weingrill}, {Werner}, {Wevers}, {Whitehead}, {Wyrzykowski}, {Yoldas}, {{\v{Z}}erjal}, {Zucker}, {Zurbach}, {Zwitter}, {Alecu}, {Allen}, {Allende Prieto}, {Amorim}, {Anglada-Escud{\'e}}, {Arsenijevic}, {Azaz}, {Balm}, {Beck}, {Bernstein},
  {Bigot}, {Bijaoui}, {Blasco}, {Bonfigli}, {Bono}, {Boudreault}, {Bressan}, {Brown}, {Brunet}, {Bunclark}, {Buonanno}, {Butkevich}, {Carret}, {Carrion}, {Chemin}, {Ch{\'e}reau}, {Corcione}, {Darmigny}, {de Boer}, {de Teodoro}, {de Zeeuw}, {Delle Luche}, {Domingues}, {Dubath}, {Fodor}, {Fr{\'e}zouls}, {Fries}, {Fustes}, {Fyfe}, {Gallardo}, {Gallegos}, {Gardiol}, {Gebran}, {Gomboc}, {G{\'o}mez}, {Grux}, {Gueguen}, {Heyrovsky}, {Hoar}, {Iannicola}, {Isasi Parache}, {Janotto}, {Joliet}, {Jonckheere}, {Keil}, {Kim}, {Klagyivik}, {Klar}, {Knude}, {Kochukhov}, {Kolka}, {Kos}, {Kutka}, {Lainey}, {LeBouquin}, {Liu}, {Loreggia}, {Makarov}, {Marseille}, {Martayan}, {Martinez-Rubi}, {Massart}, {Meynadier}, {Mignot}, {Munari}, {Nguyen}, {Nordlander}, {Ocvirk}, {O'Flaherty}, {Olias Sanz}, {Ortiz}, {Osorio}, {Oszkiewicz}, {Ouzounis}, {Palmer}, {Park}, {Pasquato}, {Peltzer}, {Peralta}, {P{\'e}turaud}, {Pieniluoma}, {Pigozzi}, {Poels}, {Prat}, {Prod'homme}, {Raison}, {Rebordao}, {Risquez}, {Rocca-Volmerange}, {Rosen},
  {Ruiz-Fuertes}, {Russo}, {Sembay}, {Serraller Vizcaino}, {Short}, {Siebert}, {Silva}, {Sinachopoulos}, {Slezak}, {Soffel}, {Sosnowska}, {Strai{\v{z}}ys}, {ter Linden}, {Terrell}, {Theil}, {Tiede}, {Troisi}, {Tsalmantza}, {Tur}, {Vaccari}, {Vachier}, {Valles}, {Van Hamme}, {Veltz}, {Virtanen}, {Wallut}, {Wichmann}, {Wilkinson}, {Ziaeepour}, \& {Zschocke}}]{GaiaCollab2016}
{Gaia Collaboration}, {Prusti}, T., {de Bruijne}, J.~H.~J., {et~al.} 2016, \aap, 595, A1, \dodoi{10.1051/0004-6361/201629272}

\bibitem[{{Gaia Collaboration} {et~al.}(2023){Gaia Collaboration}, {Vallenari}, {Brown}, {Prusti}, {de Bruijne}, {Arenou}, {Babusiaux}, {Biermann}, {Creevey}, {Ducourant}, {Evans}, {Eyer}, {Guerra}, {Hutton}, {Jordi}, {Klioner}, {Lammers}, {Lindegren}, {Luri}, {Mignard}, {Panem}, {Pourbaix}, {Randich}, {Sartoretti}, {Soubiran}, {Tanga}, {Walton}, {Bailer-Jones}, {Bastian}, {Drimmel}, {Jansen}, {Katz}, {Lattanzi}, {van Leeuwen}, {Bakker}, {Cacciari}, {Casta{\~n}eda}, {De Angeli}, {Fabricius}, {Fouesneau}, {Fr{\'e}mat}, {Galluccio}, {Guerrier}, {Heiter}, {Masana}, {Messineo}, {Mowlavi}, {Nicolas}, {Nienartowicz}, {Pailler}, {Panuzzo}, {Riclet}, {Roux}, {Seabroke}, {Sordo}, {Th{\'e}venin}, {Gracia-Abril}, {Portell}, {Teyssier}, {Altmann}, {Andrae}, {Audard}, {Bellas-Velidis}, {Benson}, {Berthier}, {Blomme}, {Burgess}, {Busonero}, {Busso}, {C{\'a}novas}, {Carry}, {Cellino}, {Cheek}, {Clementini}, {Damerdji}, {Davidson}, {de Teodoro}, {Nu{\~n}ez Campos}, {Delchambre}, {Dell'Oro}, {Esquej},
  {Fern{\'a}ndez-Hern{\'a}ndez}, {Fraile}, {Garabato}, {Garc{\'\i}a-Lario}, {Gosset}, {Haigron}, {Halbwachs}, {Hambly}, {Harrison}, {Hern{\'a}ndez}, {Hestroffer}, {Hodgkin}, {Holl}, {Jan{\ss}en}, {Jevardat de Fombelle}, {Jordan}, {Krone-Martins}, {Lanzafame}, {L{\"o}ffler}, {Marchal}, {Marrese}, {Moitinho}, {Muinonen}, {Osborne}, {Pancino}, {Pauwels}, {Recio-Blanco}, {Reyl{\'e}}, {Riello}, {Rimoldini}, {Roegiers}, {Rybizki}, {Sarro}, {Siopis}, {Smith}, {Sozzetti}, {Utrilla}, {van Leeuwen}, {Abbas}, {{\'A}brah{\'a}m}, {Abreu Aramburu}, {Aerts}, {Aguado}, {Ajaj}, {Aldea-Montero}, {Altavilla}, {{\'A}lvarez}, {Alves}, {Anders}, {Anderson}, {Anglada Varela}, {Antoja}, {Baines}, {Baker}, {Balaguer-N{\'u}{\~n}ez}, {Balbinot}, {Balog}, {Barache}, {Barbato}, {Barros}, {Barstow}, {Bartolom{\'e}}, {Bassilana}, {Bauchet}, {Becciani}, {Bellazzini}, {Berihuete}, {Bernet}, {Bertone}, {Bianchi}, {Binnenfeld}, {Blanco-Cuaresma}, {Blazere}, {Boch}, {Bombrun}, {Bossini}, {Bouquillon}, {Bragaglia}, {Bramante}, {Breedt},
  {Bressan}, {Brouillet}, {Brugaletta}, {Bucciarelli}, {Burlacu}, {Butkevich}, {Buzzi}, {Caffau}, {Cancelliere}, {Cantat-Gaudin}, {Carballo}, {Carlucci}, {Carnerero}, {Carrasco}, {Casamiquela}, {Castellani}, {Castro-Ginard}, {Chaoul}, {Charlot}, {Chemin}, {Chiaramida}, {Chiavassa}, {Chornay}, {Comoretto}, {Contursi}, {Cooper}, {Cornez}, {Cowell}, {Crifo}, {Cropper}, {Crosta}, {Crowley}, {Dafonte}, {Dapergolas}, {David}, {David}, {de Laverny}, {De Luise}, {De March}, {De Ridder}, {de Souza}, {de Torres}, {del Peloso}, {del Pozo}, {Delbo}, {Delgado}, {Delisle}, {Demouchy}, {Dharmawardena}, {Di Matteo}, {Diakite}, {Diener}, {Distefano}, {Dolding}, {Edvardsson}, {Enke}, {Fabre}, {Fabrizio}, {Faigler}, {Fedorets}, {Fernique}, {Fienga}, {Figueras}, {Fournier}, {Fouron}, {Fragkoudi}, {Gai}, {Garcia-Gutierrez}, {Garcia-Reinaldos}, {Garc{\'\i}a-Torres}, {Garofalo}, {Gavel}, {Gavras}, {Gerlach}, {Geyer}, {Giacobbe}, {Gilmore}, {Girona}, {Giuffrida}, {Gomel}, {Gomez}, {Gonz{\'a}lez-N{\'u}{\~n}ez},
  {Gonz{\'a}lez-Santamar{\'\i}a}, {Gonz{\'a}lez-Vidal}, {Granvik}, {Guillout}, {Guiraud}, {Guti{\'e}rrez-S{\'a}nchez}, {Guy}, {Hatzidimitriou}, {Hauser}, {Haywood}, {Helmer}, {Helmi}, {Sarmiento}, {Hidalgo}, {Hilger}, {H{\l}adczuk}, {Hobbs}, {Holland}, {Huckle}, {Jardine}, {Jasniewicz}, {Jean-Antoine Piccolo}, {Jim{\'e}nez-Arranz}, {Jorissen}, {Juaristi Campillo}, {Julbe}, {Karbevska}, {Kervella}, {Khanna}, {Kontizas}, {Kordopatis}, {Korn}, {K{\'o}sp{\'a}l}, {Kostrzewa-Rutkowska}, {Kruszy{\'n}ska}, {Kun}, {Laizeau}, {Lambert}, {Lanza}, {Lasne}, {Le Campion}, {Lebreton}, {Lebzelter}, {Leccia}, {Leclerc}, {Lecoeur-Taibi}, {Liao}, {Licata}, {Lindstr{\o}m}, {Lister}, {Livanou}, {Lobel}, {Lorca}, {Loup}, {Madrero Pardo}, {Magdaleno Romeo}, {Managau}, {Mann}, {Manteiga}, {Marchant}, {Marconi}, {Marcos}, {Marcos Santos}, {Mar{\'\i}n Pina}, {Marinoni}, {Marocco}, {Marshall}, {Martin Polo}, {Mart{\'\i}n-Fleitas}, {Marton}, {Mary}, {Masip}, {Massari}, {Mastrobuono-Battisti}, {Mazeh}, {McMillan}, {Messina}, {Michalik},
  {Millar}, {Mints}, {Molina}, {Molinaro}, {Moln{\'a}r}, {Monari}, {Mongui{\'o}}, {Montegriffo}, {Montero}, {Mor}, {Mora}, {Morbidelli}, {Morel}, {Morris}, {Muraveva}, {Murphy}, {Musella}, {Nagy}, {Noval}, {Oca{\~n}a}, {Ogden}, {Ordenovic}, {Osinde}, {Pagani}, {Pagano}, {Palaversa}, {Palicio}, {Pallas-Quintela}, {Panahi}, {Payne-Wardenaar}, {Pe{\~n}alosa Esteller}, {Penttil{\"a}}, {Pichon}, {Piersimoni}, {Pineau}, {Plachy}, {Plum}, {Poggio}, {Pr{\v{s}}a}, {Pulone}, {Racero}, {Ragaini}, {Rainer}, {Raiteri}, {Rambaux}, {Ramos}, {Ramos-Lerate}, {Re Fiorentin}, {Regibo}, {Richards}, {Rios Diaz}, {Ripepi}, {Riva}, {Rix}, {Rixon}, {Robichon}, {Robin}, {Robin}, {Roelens}, {Rogues}, {Rohrbasser}, {Romero-G{\'o}mez}, {Rowell}, {Royer}, {Ruz Mieres}, {Rybicki}, {Sadowski}, {S{\'a}ez N{\'u}{\~n}ez}, {Sagrist{\`a} Sell{\'e}s}, {Sahlmann}, {Salguero}, {Samaras}, {Sanchez Gimenez}, {Sanna}, {Santove{\~n}a}, {Sarasso}, {Schultheis}, {Sciacca}, {Segol}, {Segovia}, {S{\'e}gransan}, {Semeux}, {Shahaf}, {Siddiqui}, {Siebert},
  {Siltala}, {Silvelo}, {Slezak}, {Slezak}, {Smart}, {Snaith}, {Solano}, {Solitro}, {Souami}, {Souchay}, {Spagna}, {Spina}, {Spoto}, {Steele}, {Steidelm{\"u}ller}, {Stephenson}, {S{\"u}veges}, {Surdej}, {Szabados}, {Szegedi-Elek}, {Taris}, {Taylor}, {Teixeira}, {Tolomei}, {Tonello}, {Torra}, {Torra}, {Torralba Elipe}, {Trabucchi}, {Tsounis}, {Turon}, {Ulla}, {Unger}, {Vaillant}, {van Dillen}, {van Reeven}, {Vanel}, {Vecchiato}, {Viala}, {Vicente}, {Voutsinas}, {Weiler}, {Wevers}, {Wyrzykowski}, {Yoldas}, {Yvard}, {Zhao}, {Zorec}, {Zucker}, \& {Zwitter}}]{GaiaCollab2023}
{Gaia Collaboration}, {Vallenari}, A., {Brown}, A.~G.~A., {et~al.} 2023, \aap, 674, A1, \dodoi{10.1051/0004-6361/202243940}

\bibitem[{{Gilmore} {et~al.}(1989){Gilmore}, {Wyse}, \& {Kuijken}}]{Gilmore1989}
{Gilmore}, G., {Wyse}, R. F.~G., \& {Kuijken}, K. 1989, \araa, 27, 555, \dodoi{10.1146/annurev.aa.27.090189.003011}

\bibitem[{{Hasselquist} {et~al.}(2020){Hasselquist}, {Zasowski}, {Feuillet}, {Schultheis}, {Nataf}, {Anguiano}, {Beaton}, {Beers}, {Cohen}, {Cunha}, {Fern{\'a}ndez-Trincado}, {Garc{\'\i}a-Hern{\'a}ndez}, {Geisler}, {Holtzman}, {Johnson}, {Lane}, {Majewski}, {Moni Bidin}, {Nitschelm}, {Roman-Lopes}, {Schiavon}, {Smith}, \& {Sobeck}}]{Hasselquist2020}
{Hasselquist}, S., {Zasowski}, G., {Feuillet}, D.~K., {et~al.} 2020, \apj, 901, 109, \dodoi{10.3847/1538-4357/abaeee}

\bibitem[{{Hayden} {et~al.}(2015){Hayden}, {Bovy}, {Holtzman}, {Nidever}, {Bird}, {Weinberg}, {Andrews}, {Majewski}, {Allende Prieto}, {Anders}, {Beers}, {Bizyaev}, {Chiappini}, {Cunha}, {Frinchaboy}, {Garc{\'\i}a-Her{\'n}andez}, {Garc{\'\i}a P{\'e}rez}, {Girardi}, {Harding}, {Hearty}, {Johnson}, {M{\'e}sz{\'a}ros}, {Minchev}, {O'Connell}, {Pan}, {Robin}, {Schiavon}, {Schneider}, {Schultheis}, {Shetrone}, {Skrutskie}, {Steinmetz}, {Smith}, {Wilson}, {Zamora}, \& {Zasowski}}]{Hayden2015}
{Hayden}, M.~R., {Bovy}, J., {Holtzman}, J.~A., {et~al.} 2015, ApJ, 808, 132, \dodoi{10.1088/0004-637X/808/2/132}

\bibitem[{{Horta} {et~al.}(2024){Horta}, {Petersen}, \& {Pe{\~n}arrubia}}]{Horta2024}
{Horta}, D., {Petersen}, M.~S., \& {Pe{\~n}arrubia}, J. 2024, arXiv e-prints, arXiv:2402.07986, \dodoi{10.48550/arXiv.2402.07986}

\bibitem[{{Johnson} {et~al.}(2022){Johnson}, {Rich}, {Simion}, {Young}, {Clarkson}, {Pilachowski}, {Michael}, {Marchetti}, {Soto}, {Kunder}, {Koch-Hansen}, {Katherina Vivas}, {Joyce}, {Shen}, \& {Osmond}}]{Johnson2022}
{Johnson}, C.~I., {Rich}, R.~M., {Simion}, I.~T., {et~al.} 2022, \mnras, 515, 1469, \dodoi{10.1093/mnras/stac1840}

\bibitem[{{Kollmeier} {et~al.}(2017){Kollmeier}, {Zasowski}, {Rix}, {Johns}, {Anderson}, {Drory}, {Johnson}, {Pogge}, {Bird}, {Blanc}, {Brownstein}, {Crane}, {De Lee}, {Klaene}, {Kreckel}, {MacDonald}, {Merloni}, {Ness}, {O'Brien}, {Sanchez-Gallego}, {Sayres}, {Shen}, {Thakar}, {Tkachenko}, {Aerts}, {Blanton}, {Eisenstein}, {Holtzman}, {Maoz}, {Nandra}, {Rockosi}, {Weinberg}, {Bovy}, {Casey}, {Chaname}, {Clerc}, {Conroy}, {Eracleous}, {G{\"a}nsicke}, {Hekker}, {Horne}, {Kauffmann}, {McQuinn}, {Pellegrini}, {Schinnerer}, {Schlafly}, {Schwope}, {Seibert}, {Teske}, \& {van Saders}}]{Kollmeier2017}
{Kollmeier}, J.~A., {Zasowski}, G., {Rix}, H.-W., {et~al.} 2017, ArXiv e-prints.
\newblock \doarXiv{1711.03234}

\bibitem[{{Li} {et~al.}(2024){Li}, {Wong}, {Hogg}, {Rix}, \& {Chandra}}]{Li2023}
{Li}, J., {Wong}, K. W.~K., {Hogg}, D.~W., {Rix}, H.-W., \& {Chandra}, V. 2024, \apjs, 272, 2, \dodoi{10.3847/1538-4365/ad2b4d}

\bibitem[{{Lian} {et~al.}(2020){Lian}, {Zasowski}, {Hasselquist}, {Nataf}, {Thomas}, {Moni Bidin}, {Fern{\'a}ndez-Trincado}, {Garcia-Hernandez}, {Lane}, {Majewski}, {Roman-Lopes}, \& {Schultheis}}]{Lian2020}
{Lian}, J., {Zasowski}, G., {Hasselquist}, S., {et~al.} 2020, \mnras, 497, 3557, \dodoi{10.1093/mnras/staa2205}

\bibitem[{{Majewski} {et~al.}(2017){Majewski}, {Schiavon}, {Frinchaboy}, {Allende Prieto}, {Barkhouser}, {Bizyaev}, {Blank}, {Brunner}, {Burton}, {Carrera}, {Chojnowski}, {Cunha}, {Epstein}, {Fitzgerald}, {Garc{\'{\i}}a P{\'e}rez}, {Hearty}, {Henderson}, {Holtzman}, {Johnson}, {Lam}, {Lawler}, {Maseman}, {M{\'e}sz{\'a}ros}, {Nelson}, {Nguyen}, {Nidever}, {Pinsonneault}, {Shetrone}, {Smee}, {Smith}, {Stolberg}, {Skrutskie}, {Walker}, {Wilson}, {Zasowski}, {Anders}, {Basu}, {Beland}, {Blanton}, {Bovy}, {Brownstein}, {Carlberg}, {Chaplin}, {Chiappini}, {Eisenstein}, {Elsworth}, {Feuillet}, {Fleming}, {Galbraith-Frew}, {Garc{\'{\i}}a}, {Garc{\'{\i}}a-Hern{\'a}ndez}, {Gillespie}, {Girardi}, {Gunn}, {Hasselquist}, {Hayden}, {Hekker}, {Ivans}, {Kinemuchi}, {Klaene}, {Mahadevan}, {Mathur}, {Mosser}, {Muna}, {Munn}, {Nichol}, {O'Connell}, {Parejko}, {Robin}, {Rocha-Pinto}, {Schultheis}, {Serenelli}, {Shane}, {Silva Aguirre}, {Sobeck}, {Thompson}, {Troup}, {Weinberg}, \& {Zamora}}]{Majewski2017}
{Majewski}, S.~R., {Schiavon}, R.~P., {Frinchaboy}, P.~M., {et~al.} 2017, \aj, 154, 94, \dodoi{10.3847/1538-3881/aa784d}

\bibitem[{{Minchev} {et~al.}(2013){Minchev}, {Chiappini}, \& {Martig}}]{Minchev2013}
{Minchev}, I., {Chiappini}, C., \& {Martig}, M. 2013, \aap, 558, A9, \dodoi{10.1051/0004-6361/201220189}

\bibitem[{{Naidu} {et~al.}(2020){Naidu}, {Conroy}, {Bonaca}, {Johnson}, {Ting}, {Caldwell}, {Zaritsky}, \& {Cargile}}]{Naidu2020}
{Naidu}, R.~P., {Conroy}, C., {Bonaca}, A., {et~al.} 2020, \apj, 901, 48, \dodoi{10.3847/1538-4357/abaef4}

\bibitem[{{Nelson} {et~al.}(2019){Nelson}, {Springel}, {Pillepich}, {Rodriguez-Gomez}, {Torrey}, {Genel}, {Vogelsberger}, {Pakmor}, {Marinacci}, {Weinberger}, {Kelley}, {Lovell}, {Diemer}, \& {Hernquist}}]{Nelson2019}
{Nelson}, D., {Springel}, V., {Pillepich}, A., {et~al.} 2019, Computational Astrophysics and Cosmology, 6, 2, \dodoi{10.1186/s40668-019-0028-x}

\bibitem[{{Nepal} {et~al.}(2024){Nepal}, {Chiappini}, {Guiglion}, {Steinmetz}, {P{\'e}rez-Villegas}, {Queiroz}, {Miglio}, {Dohme}, \& {Khalatyan}}]{Nepal24}
{Nepal}, S., {Chiappini}, C., {Guiglion}, G., {et~al.} 2024, \aap, 681, L8, \dodoi{10.1051/0004-6361/202348365}

\bibitem[{{Ness} \& {Freeman}(2016)}]{Ness2016b}
{Ness}, M., \& {Freeman}, K. 2016, PASA, 33, e022, \dodoi{10.1017/pasa.2015.51}

\bibitem[{{Ness} {et~al.}(2016){Ness}, {Zasowski}, {Johnson}, {Athanassoula}, {Majewski}, {Garc{\'{\i}}a P{\'e}rez}, {Bird}, {Nidever}, {Schneider}, {Sobeck}, {Frinchaboy}, {Pan}, {Bizyaev}, {Oravetz}, \& {Simmons}}]{Ness2016a}
{Ness}, M., {Zasowski}, G., {Johnson}, J.~A., {et~al.} 2016, \apj, 819, 2, \dodoi{10.3847/0004-637X/819/1/2}

\bibitem[{{Pillepich} {et~al.}(2019){Pillepich}, {Nelson}, {Springel}, {Pakmor}, {Torrey}, {Weinberger}, {Vogelsberger}, {Marinacci}, {Genel}, {van der Wel}, \& {Hernquist}}]{Pillepich2019}
{Pillepich}, A., {Nelson}, D., {Springel}, V., {et~al.} 2019, \mnras, 490, 3196, \dodoi{10.1093/mnras/stz2338}

\bibitem[{{Pillepich} {et~al.}(2023){Pillepich}, {Sotillo-Ramos}, {Ramesh}, {Nelson}, {Engler}, {Rodriguez-Gomez}, {Fournier}, {Donnari}, {Springel}, \& {Hernquist}}]{Pillepich2023}
{Pillepich}, A., {Sotillo-Ramos}, D., {Ramesh}, R., {et~al.} 2023, arXiv e-prints, arXiv:2303.16217, \dodoi{10.48550/arXiv.2303.16217}

\bibitem[{{Queiroz} {et~al.}(2021){Queiroz}, {Chiappini}, {Perez-Villegas}, {Khalatyan}, {Anders}, {Barbuy}, {Santiago}, {Steinmetz}, {Cunha}, {Schultheis}, {Majewski}, {Minchev}, {Minniti}, {Beaton}, {Cohen}, {da Costa}, {Fern{\'a}ndez-Trincado}, {Garcia-Hern{\'a}ndez}, {Geisler}, {Hasselquist}, {Lane}, {Nitschelm}, {Rojas-Arriagada}, {Roman-Lopes}, {Smith}, \& {Zasowski}}]{Queiroz21}
{Queiroz}, A.~B.~A., {Chiappini}, C., {Perez-Villegas}, A., {et~al.} 2021, \aap, 656, A156, \dodoi{10.1051/0004-6361/202039030}

\bibitem[{{Rix} \& {Bovy}(2013)}]{Rix2013}
{Rix}, H.-W., \& {Bovy}, J. 2013, A\&ARv, 21, 61, \dodoi{10.1007/s00159-013-0061-8}

\bibitem[{{Rix} {et~al.}(2022){Rix}, {Chandra}, {Andrae}, {Price-Whelan}, {Weinberg}, {Conroy}, {Fouesneau}, {Hogg}, {De Angeli}, {Naidu}, {Xiang}, \& {Ruz-Mieres}}]{Rix2022}
{Rix}, H.-W., {Chandra}, V., {Andrae}, R., {et~al.} 2022, \apj, 941, 45, \dodoi{10.3847/1538-4357/ac9e01}

\bibitem[{{Rojas-Arriagada} {et~al.}(2014){Rojas-Arriagada}, {Recio-Blanco}, {Hill}, {de Laverny}, {Schultheis}, {Babusiaux}, {Zoccali}, {Minniti}, {Gonzalez}, {Feltzing}, {Gilmore}, {Randich}, {Vallenari}, {Alfaro}, {Bensby}, {Bragaglia}, {Flaccomio}, {Lanzafame}, {Pancino}, {Smiljanic}, {Bergemann}, {Costado}, {Damiani}, {Hourihane}, {Jofr{\'e}}, {Lardo}, {Magrini}, {Maiorca}, {Morbidelli}, {Sbordone}, {Worley}, {Zaggia}, \& {Wyse}}]{Rojas-Arriagada14}
{Rojas-Arriagada}, A., {Recio-Blanco}, A., {Hill}, V., {et~al.} 2014, \aap, 569, A103, \dodoi{10.1051/0004-6361/201424121}

\bibitem[{{Rojas-Arriagada} {et~al.}(2020){Rojas-Arriagada}, {Zasowski}, {Schultheis}, {Zoccali}, {Hasselquist}, {Chiappini}, {Cohen}, {Cunha}, {Fern{\'a}ndez-Trincado}, {Fragkoudi}, {Garc{\'\i}a-Hern{\'a}ndez}, {Geisler}, {Gran}, {Lian}, {Majewski}, {Minniti}, {Monachesi}, {Nitschelm}, \& {Queiroz}}]{Rojas2020}
{Rojas-Arriagada}, A., {Zasowski}, G., {Schultheis}, M., {et~al.} 2020, \mnras, 499, 1037, \dodoi{10.1093/mnras/staa2807}

\bibitem[{{Semenov} {et~al.}(2024){Semenov}, {Conroy}, {Chandra}, {Hernquist}, \& {Nelson}}]{Semenov2024}
{Semenov}, V.~A., {Conroy}, C., {Chandra}, V., {Hernquist}, L., \& {Nelson}, D. 2024, \apj, 962, 84, \dodoi{10.3847/1538-4357/ad150a}

\bibitem[{{Sestito} {et~al.}(2020){Sestito}, {Martin}, {Starkenburg}, {Arentsen}, {Ibata}, {Longeard}, {Kielty}, {Youakim}, {Venn}, {Aguado}, {Carlberg}, {Gonz{\'a}lez Hern{\'a}ndez}, {Hill}, {Jablonka}, {Kordopatis}, {Malhan}, {Navarro}, {S{\'a}nchez-Janssen}, {Thomas}, {Tolstoy}, {Wilson}, {Palicio}, {Bialek}, {Garcia-Dias}, {Lucchesi}, {North}, {Osorio}, {Patrick}, \& {Peralta de Arriba}}]{Sestito2020}
{Sestito}, F., {Martin}, N.~F., {Starkenburg}, E., {et~al.} 2020, \mnras, 497, L7, \dodoi{10.1093/mnrasl/slaa022}

\bibitem[{{Soubiran} {et~al.}(2022){Soubiran}, {Brouillet}, \& {Casamiquela}}]{Soubiran2022}
{Soubiran}, C., {Brouillet}, N., \& {Casamiquela}, L. 2022, \aap, 663, A4, \dodoi{10.1051/0004-6361/202142409}

\bibitem[{{Valenti} {et~al.}(2018){Valenti}, {Zoccali}, {Mucciarelli}, {Gonzalez}, {Surot}, {Minniti}, {Rejkuba}, {Pasquini}, {Fiorentino}, {Bono}, {Rich}, \& {Soto}}]{Valenti2018}
{Valenti}, E., {Zoccali}, M., {Mucciarelli}, A., {et~al.} 2018, \aap, 616, A83, \dodoi{10.1051/0004-6361/201832905}

\bibitem[{{Weinberg} {et~al.}(2017){Weinberg}, {Andrews}, \& {Freudenburg}}]{Weinberg2017}
{Weinberg}, D.~H., {Andrews}, B.~H., \& {Freudenburg}, J. 2017, \apj, 837, 183, \dodoi{10.3847/1538-4357/837/2/183}

\bibitem[{{Weinberg} {et~al.}(2019){Weinberg}, {Holtzman}, {Hasselquist}, {Bird}, {Johnson}, {Shetrone}, {Sobeck}, {Allende Prieto}, {Bizyaev}, {Carrera}, {Cohen}, {Cunha}, {Ebelke}, {Fernandez-Trincado}, {Garc{\'\i}a-Hern{\'a}ndez}, {Hayes}, {J{\"o}nsson}, {Lane}, {Majewski}, {Malanushenko}, {M{\'e}sz{\'a}ros}, {Nidever}, {Nitschelm}, {Pan}, {Rix}, {Rybizki}, {Schiavon}, {Schneider}, {Wilson}, \& {Zamora}}]{Weinberg2019}
{Weinberg}, D.~H., {Holtzman}, J.~A., {Hasselquist}, S., {et~al.} 2019, ApJ, 874, 102, \dodoi{10.3847/1538-4357/ab07c7}

\bibitem[{{Xiang} \& {Rix}(2022)}]{Xiang2022}
{Xiang}, M., \& {Rix}, H.-W. 2022, Natur, 603, 599, \dodoi{10.1038/s41586-022-04496-5}

\bibitem[{{Youakim} {et~al.}(2020){Youakim}, {Starkenburg}, {Martin}, {Matijevi{\v{c}}}, {Aguado}, {Allende Prieto}, {Arentsen}, {Bonifacio}, {Carlberg}, {Gonz{\'a}lez Hern{\'a}ndez}, {Hill}, {Kordopatis}, {Lardo}, {Navarro}, {Jablonka}, {S{\'a}nchez Janssen}, {Sestito}, {Thomas}, \& {Venn}}]{Youakim2020}
{Youakim}, K., {Starkenburg}, E., {Martin}, N.~F., {et~al.} 2020, MNRAS, 492, 4986, \dodoi{10.1093/mnras/stz3619}

\bibitem[{{Zana} {et~al.}(2022){Zana}, {Lupi}, {Bonetti}, {Dotti}, {Rosas-Guevara}, {Izquierdo-Villalba}, {Bonoli}, {Hernquist}, \& {Nelson}}]{Zana2022}
{Zana}, T., {Lupi}, A., {Bonetti}, M., {et~al.} 2022, \mnras, 515, 1524, \dodoi{10.1093/mnras/stac1708}

\bibitem[{{Zasowski} {et~al.}(2016){Zasowski}, {Ness}, {Garc{\'\i}a P{\'e}rez}, {Martinez-Valpuesta}, {Johnson}, \& {Majewski}}]{Zasowski2016}
{Zasowski}, G., {Ness}, M.~K., {Garc{\'\i}a P{\'e}rez}, A.~E., {et~al.} 2016, \apj, 832, 132, \dodoi{10.3847/0004-637X/832/2/132}

\bibitem[{{Zasowski} {et~al.}(2019){Zasowski}, {Schultheis}, {Hasselquist}, {Cunha}, {Sobeck}, {Johnson}, {Rojas-Arriagada}, {Majewski}, {Andrews}, {J{\"o}nsson}, {Beers}, {Chojnowski}, {Frinchaboy}, {Holtzman}, {Minniti}, {Nidever}, \& {Nitschelm}}]{Zasowski2019}
{Zasowski}, G., {Schultheis}, M., {Hasselquist}, S., {et~al.} 2019, \apj, 870, 138, \dodoi{10.3847/1538-4357/aaeff4}

\bibitem[{{Zhang} {et~al.}(2023){Zhang}, {Green}, \& {Rix}}]{Zhang2023}
{Zhang}, X., {Green}, G.~M., \& {Rix}, H.-W. 2023, MNRAS, 524, 1855, \dodoi{10.1093/mnras/stad1941}

\end{thebibliography}
\bibliographystyle{aasjournal}



\end{document}